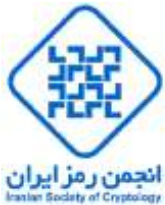

# منادی

## موضوع خاص این شماره: نهان نگاری (۱)

**دبیر مدعو:** دکتر شاهرخ قائم مقامی

## در این شماره:

## سخن سردبیر

این شماره و شماره بعد از نشریه منادی با موضوع خاص "نهان نگاری" به خوانندگان محترم تقدیم می‌شود. قطعا بدون حمایت و نظارت همکار محترم جناب آقای دکتر شاهرخ قائم‌مقامی که به عنوان دبیر مدعو این شماره قبول زحمت نمودند، ارائه این مطالب و مقالات که انشاالله مورد توجه علاقمندان قرار خواهد گرفت، میسر نمی‌شد. در این مجموعه جمعا شش مقاله به حضور خوانندگان محترم ارائه می‌شود که سه مقاله آن در این شماره و سه مقاله دیگر در شماره بعدی خواهند بود.

ضمنا از طرف هیئت تحریر منادی از "جناب آقای مهندس داود فرج‌زاده" نیز که در صفحه آرایی جدید منادی همکاری نموده و موجب ارتقاء کیفی این نشریه شدند صمیمانه سپاسگزاریم و مجددا از کلیه همکاران و اعضای محترم انجمن و علاقمندان جهت همکاری در ارائه هر چه بهتر این نشریه درخواست مساعدت و مشارکت فعال داریم.

سر دبیر

# نهان‌نگاری بر اساس تجزیه تنک بلوک‌های تصویر

سوده آهنی

دانشکده مهندسی برق – دانشگاه صنعتی شریف

s_ahani@ee.sharif.edu

## چکیده

نمایش تُنُک سیگنال‌ها برای فشرده سازی و تحلیل محتوا (ساختار یا معنای سیگنال) به کار برده می‌شود و در حوزه پردازش تصویر مورد توجه است. تمایل برای داشتن یک نمایش فشرده در پردازش سیگنال، در کاربردهایی مانند کدینگ، حذف نویز و آنالیز سیگنال منجر به نیاز به جستجوی بسط تنک سیگنال شده است. لیکن ایده‌ی استفاده از تجزیه تنک تا کنون در زمینه نهان‌نگاری و نشان‌گذاری استفاده نشده است. در این مقاله الگوریتم نهان‌نگاری نوینی بر مبنای تجزیه تنک بلوک‌های تصویر بر روی دیکشنریِ بدست آمده از تصویر پوشش، ارائه شده است. امکان استفاده از این روش بر روی سایر انواع سیگنال‌های پوشش و همچنین کاربرد نشان‌گذاری نیز وجود دارد.

واژه های کلیدی : نهان‌نگاری، تجزیه تنک، الگوریتم KSVD، تخمین دیکشنری .

## 1 – مقدمه

تکنیک های مخفی‌سازی اطلاعات شامل رمزنگاری، نشان‌گذاری و نهان‌نگاری است. در رمزنگاری پیام توسط یک کلید رمز می شود.

در این حالت وجود پیام آشکار است و هدف این است که تنها گیرنده‌ای که از کلید مطلع است بتواند به پیام دسترسی داشته باشد. نشان‌گذاری روش درج داده به منظور حفظ مالکیت محصولات دیجیتال است. در نشان‌گذاری یک رشته بیت توسط یک کلید درون سیگنال پوشش به گونه‌ای درج می‌شود که حذفش بدون داشتن اطلاعات اضافی همچون کلید امکان‌پذیر نباشد. در نهایت، نهان‌نگاری به معنای درج بیت‌های پیام محرمانه در سیگنال پوشش یا حامل به گونه ای است که وجود پیام از نظر ناظر به کانال و گیرنده‌های غیر مجاز مخفی باشد. به دلیل زمینه‌های اقتصادی و توجه دولت ها به این زمینه، تحقیقات در زمینه‌ی مخف سازی اطلاعات در دهه های اخیر رشد روز افزونی داشته است. در این مقاله یک الگوریتم نهان‌نگاری معرفی شده است که از الگوریتم آموزش دیکشنری KSVD [1] استفاده می‌کند و دیکشنری مورد نیاز برای نمایش تنک بلوک های تصویر را بدست می آورد و بیت های پیام را درون ضرایب غیر صفر بسط تنک بلوک‌ها درج می کند. در ادامه ابتدا به بیان مقدمه‌ای در زمینه‌ی بسط تنک سیگنال و الگوریتم تخمین دیکشنری KSVD [1] می‌پردازیم. سپس الگوریتم نهان‌نگاری پیشنهادی را بیان کرده و در نهایت به بیان نتایج بدست‌آمده و نتیجه‌گیری خواهیم پرداخت.

---

[1] Dictionary learning

# نهان‌نگاری بر اساس تجزیه تنک بلوک‌های تصویر

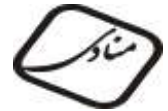

## 2- تجزیه تنک سیگنال

یافتن جواب‌های تنک دستگاه‌های معادلات خطی «کمتر از حد معین»[1] یعنی دستگاه‌هایی که تعداد مجهولات آن‌ها بیشتر از تعداد معادلات است، یکی از مسائلی است که در سال‌های اخیر مورد توجه فراوان محققین قرار گرفته است [2-5]. نمایش تنک سیگنال‌ها برای فشرده سازی و تحلیل محتوا (ساختار یا معنای سیگنال) به کار برده می‌شود و در حوزه پردازش تصویر مورد توجه است. . تمایل برای داشتن یک نمایش فشرده در پردازش سیگنال، در کاربردهایی مانند کدینگ، حذف نویز و آنالیز سیگنال منجر به نیاز به جستجوی بسط تنک سیگنال شده است. تنک بودن یک بردار بدین معناست که تعداد درایه‌های غیر صفر آن نسبت به بُعد سیگنال اندک است. هدف از مسأله‌ی «تجزیه‌ی اتمی»[2] یا «تجزیه به مولفه‌های تنک»[3] یافتن نمایشی تنک برای سیگنال $x$، $m \times 1$ به شکل ترکیب خطی از $n$ بردار (سیگنال) از پیش تعیین شده‌ی $\{a_i\}_{i=1}^n$ است. بردار $x$ سیگنال مخلوط نامیده می‌شود. بردارهای $a_i$، $1 \leq i \leq n$ را اتم و ماتریس $A$ که ستون‌هایش بردارهای $a_i$ هستند، را ماتریس دیکشنری می‌نامیم. این اصطلاحات نخستین بار توسط Mallat و zhang در [6] معرفی شدند و بعدها محبوبیت فراوانی پیدا کردند. به بیان ریاضی، مسأله‌ی تجزیه‌ی اتمی را می‌توان به شکل زیر نوشت:

$$\mathbf{x} = \sum_{i=1}^{n} \alpha_i \mathbf{a}_i = \mathbf{A}\boldsymbol{\alpha} \quad ; \boldsymbol{\alpha} = [\alpha_1, ..., \alpha_n]^T \quad (1)$$

در رابطه‌ی (1)، $\mathbf{A} \times [\mathbf{a}_1, ..., \mathbf{a}_n]$ ماتریس دیکشنری $m \times n$ و $\boldsymbol{\alpha} = [\alpha_1, ..., \alpha_n]^T$ بردار ضرایب $n \times 1$ است. بردار $\boldsymbol{\alpha}$ را بردار ضرایب بسط تنک می‌نامیم. دیکشنری $\mathbf{A}$ را با $n$ اتم که $n > m$ است «فوقِ کامل»[4] می‌نامند. هدف از مسأله‌ی «تجزیه‌ی اتمی»، یافتن بردار ضرایب بسط تنک است.

فرض دانستن ماتریس $\mathbf{A}$ به تنهایی برای یافتن بردار ضرایب $\boldsymbol{\alpha}$ کافی نیست؛ در واقع با فرض دانستن $\mathbf{A}$، برای بازیابی بردار منبع $\boldsymbol{\alpha}$ نیاز به حل دستگاه «کمتر از حد معین» $\mathbf{x} = \mathbf{A}\boldsymbol{\alpha}$ داریم که در حالت کلی تعداد نامتناهی جواب دارد. اما با توجه به این مطلب که در اکثر کاربردهای «تجزیه‌ی فوقِ کامل» نیاز به تنک‌ترین پاسخ رابطه‌ی (1) داریم، می‌توان پاسخ یکتای مسأله را بدست آورد. تنک‌بودن بردار منبع بدین معنا است که اکثر مولفه‌های بردار $\boldsymbol{\alpha}$ صفر هستند و تنها تعداد اندکی از درایه‌ها غیر صفر هستند.

تعداد جواب‌های دستگاه معادلات خطیِ کمتر از حد معین، در صورت وجود نامتناهی است. با این وجود، افزودن شرط‌هایی مانند تنک‌بودن پاسخ، که در بسیاری کاربردها فرض مفیدی است، منجر به یافتن پاسخ یکتای مساله خواهد شد. اثبات می‌شود که تحت شرایط خاصی از ماتریسِ دیکشنری، تنک‌ترین جواب یکتاست [7-8]. تنک بودن یک بردار بدین معناست که تعداد عناصر غیرصفر بردار نسبت به تعداد کل درایه‌ها به اندازه کافی کوچک باشد و به عبارتی نرمِ $\ell_0$ آن کوچک باشد.

در برخی کاربردها دیکشنری فوق کامل $\mathbf{A}$ موجود است، اما در دیگر کاربردها نیاز به ساخت دیکشنری متناسب با داده‌های موجود است. یکی از الگوریتم‌های وفقی برای ساخت دیکشنری از روی سیگنال‌های مخلوط موجود KSVD است. در ادامه به بیان مختصر روش تخمین دیکشنری KSVD با استفاده از بردارهای مخلوط موجود، به منظور نمایش تنک این بردارها بر حسب دیکشنری به دست آمده، می‌پردازیم.

### 2-1- KSVD روش تخمین دیکشنری

در الگوریتم KSVD دیکشنری فوق کامل از طریق حداقل‌سازی بر اساس رابطه زیر حاصل می‌شود.

$$min_{D,X} \{\|Y - DX\|_F^2\} \text{ s.t. } \forall i, \|x_i\|_0 \leq T_0$$
(2)

که ماتریس داده‌ها نامیده می‌شود و هر ستون آن یک بردار سیگنال مخلوط است. D ماتریس دیکشنری است که قصد تخمین زدنش را داریم و X ماتریس بسط تنک است که هر ستونش ($x_i$) بسط تنک

---
[1] Under Determined
[2] Atomic Decomposition
[3] Sparse Component Analysis (SCA)
[4] Over Complete

ستون متناظر از Y بر روی دیکشنری D است. در این روش با اعمال شرط بر روی نرم صفر بسط تنک، تخمین اتم‌های دیکشنری و بسط تنک سیگنال همزمان محاسبه می‌شود. در این الگوریتم بسط تنک سیگنال بر روی تخمین دیکشنری توسط روش MP[1] محاسبه می‌شود. در این روش شرط قرارداده‌شده بر روی نرم $\ell_0$ بسط تنک، شرطی است که موجب محدودیت می‌شود و با تغییر این شرط می‌توان دیکشنری‌های متفاوتی بدست آورد. اتم‌های دیکشنری توسط تجزیه‌ی SVD خطای تخمین بدست می‌آیند و در نهایت خروجی‌های این الگوریتم دیکشنری فوق کامل تخمین‌زده‌شده بر اساس بردارهای مخلوط ورودی و بسط تنک بردارهای مخلوط بر روی این دیکشنری هستند.

## ۳- الگوریتم درج بر مبنای تجزیه تنک بلوک های تصویر

در زمینه نهان‌نگاری تا کنون از روش‌های مختلفی استفاده شده است که در برخی از این روش‌ها درج مستقیماً در فضای مکانی و در برخی روش‌ها در حوزه تبدیل صورت می‌گیرد. منظور از درج در فضای مکانی این است که بیت‌های پیام در نقاط مشخصی از تصویر درج می‌شوند و در مقابل مقصود از درج در فضای تبدیل این است که فرآیند درج بر روی تبدیل یافته سیگنال به طور مثال تبدیل فوریه تصویر صورت می‌پذیرد.

یکی از روش‌های جدید که به دلیل ظرفیت کم برای نشان‌گذاری استفاده می شود استفاده از تبدیل SVD و درج در بردار مقادیر تکین است. در این مقاله یک روش نهان‌نگاری با استفاده از تجزیه تنک سیگنال ارائه می‌شود که احتمالاً اولین روشی است که ایده‌ی استفاده از تجزیه تنک سیگنال به منظور درج داده پنهان را مطرح می‌کند.

شکل ۱ الگوریتم نهان نگاری ارائه شده در این مقاله را نشان می‌دهد.

---

۱- شکستن تصویر به بلوک های $n \times n$.

۲- تبدیل تصویر بلوک ها به بردار های $n^2 \times 1$ و چیدن بردارهای $x_{n^2 \times 1}$ به عنوان ستون های ماتریس داده $X_{n^2 \times J}$.

۳- تعیین تعداد اتم های دیکشنری (تعداد ستون‌های دیکشنری) k و تعداد ضرایب غیر صفر (m) در بسط تنک داده.

۴- استفاده از الگوریتم KSVD با ورودی های X، m و k و بدست آوردن ماتریس دیکشنری تخمین زده شده‌ی D و ماتریس Ψ که هر ستون آن ضرایب بسط تنک ستون متناظر از ماتریس X بر روی دیکشنری D است. به عبارت دیگر:

$$X_{n^2 \times J} = D_{n^2 \times k} \times \Psi_{k \times J}$$

۵- یافتن درایه های غیر صفر ماتریس Ψ.

۶- برای هر درایه‌ی غیر صفر $\varphi(i,j)$، ۴ بیت کم ارزش قسمت اعشاری $\varphi(i,j)$ را با بیت k ام پیام ($P_k$) جایگزین می‌کنیم و ماتریس ضرایب جدید $\widetilde{\Psi}$ را می‌سازیم.

۷- با استفاده از رابطه زیر ماتریس $\widetilde{X}$ را بدست می‌آوریم:

$$\widetilde{X_{n^2 \times J}} = D_{n^2 \times k} \times \widetilde{\Psi_{k \times J}}$$

۹- انجام عکس عمل برداری‌سازی انجام‌شده در مرحله ۳ و ساخت بلوک های تصویر نهان‌نگاری‌شده.

۱۰- ساخت تصویر نهان‌نگاری‌شده نهایی با استفاده از بلوک‌های بدست‌آمده در مرحله ۹.

شکل ۱: مراحل درج الگوریتم نهان‌نگاری پیشنهادی.

همانطور که در شکل ۱ بیان شده است، در ابتدا تصویر به بلوک‌های $n \times n$ شکسته می‌شود سپس با زیر هم قراردادن ستون‌های هر بلوک، بردار متناظر با آن بلوک را بدست می‌آوریم. تعیین اندازه

---

[1] Matching Pursuit

# نهان‌نگاری بر اساس تجزیه تنک بلوک‌های تصویر

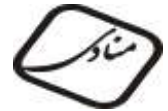

بلوک‌ها در ظرفیت نهان‌نگاری موثر است زیرا اتم‌های دیکشنری حاصل از الگوریتم KSVD نیز $n^2$ بعدی خواهند بود و طبق قضیه مطرح‌شده در [۸] شرط یکتایی تنک‌ترین پاسخ این است که تعداد درایه‌های غیر صفر بردار ضرایب (k) کمتر از نصف بعد اتم‌های دیکشنری باشد و به عبارتی می‌بایست $k < \dfrac{n^2}{2}$ باشد.

از آن جا که درج بیت‌های پیام در ضرایب غیر صفر بسط تنک $\varphi(i,j)$ بردار بلوک‌های تصویر ($x_i$) انجام می‌شود، ظرفیت درج روش پیشنهادی برابر $J \times K$ خواهد بود، که J تعداد بلوک‌های تصویر و K تعداد ضرایب غیرصفر بسط تنک بردار متناظر با هر بلوک بر روی دیکشنری بدست آمده است. بنابر این تعیین اندازه بلوک‌ها بر تعداد بلوک‌ها و حد مجاز انتخاب k و به عبارتی بر ظرفیت نهان‌نگاری تاثیر می‌گذارد. با استفاده از روابط مطرح‌شده حد بالای ظرفیت الگوریتم پیشنهادی $J \times \left(\dfrac{n^2}{2}\right)$ است.

پس از یافتن دیکشنری مناسب برای نمایش تنک بلوک‌های تصویر، می‌توان از ضرایب غیر صفر بسط تنک برای درج بیت‌های پیام مخفی استفاده کرد. بدین ترتیب ضرایب بسط تنک بلوک‌ها تغییر جزئی دارد و بر شفافیت تصویر نهان‌نگاری‌شده حاصل و هیستوگرام تصویر تاثیر جزئی خواهد داشت.

الگوریتم استخراج داده‌های درج‌شده در شکل ۲ نشان داده شده است. کلید الگوریتم، دیکشنری بدست آمده در مرحله درج است که البته می‌توان به جای استفاده از دیکشنری به عنوان کلید، پارامترهای مورد استفاده در ساخت دیکشنری توسط الگوریتم KSVD را به عنوان کلید استفاده کرد. در اینجا فرض می‌کنیم دیکشنری به عنوان کلید در دسترس است.

## ٤- نتایج تجربی

الگوریتم پیشنهادی توسط نرم افزار Matlab شبیه سازی شده است. تصویر‌های پوشش با ابعاد 256*256 به عنوان تصاویر پوشش انتخاب شده‌اند. در ابتدا تصویر به بلوک های 8*8 شکسته شده و سپس طبق الگوریتم بیان شده در شکل ۱، بردار متناظر با بلوک‌های تصویر ساخته می‌شود و از این بردارها به عنوان ستون‌های ماتریس داده‌ی X استفاده می‌شود.

برای ساخت دیکشنری توسط الگوریتم KSVD تعداد اتم‌ها (ستون‌های دیکشنری) و تعداد ضرایب غیر صفر بسط تنک (k) را تعیین کرده و با استفاده از ماتریس داده‌ی حاصل از برداری‌کردن بلوک‌های تصویر، دیکشنری را بدست می‌آوریم.

---

۱- شکستن تصویر نهان نگاری شده به بلوک های $n \times n$.

۲- تبدیل تصویر بلوک ها به بردار های $n^2 \times 1$، چیدن بردارهای $\hat{x}_{n^2 \times 1}$ به عنوان ستون های ماتریس داده $\hat{X}_{n^2 \times J}$.

۳- یافتن بسط تنک ماتریس $\hat{X}_{n^2 \times J}$ بر روی دیکشنری (کلید) و یافتن ماتریس ضرایب بسط تنک توسط الگوریتم تجزیه تنک OMP [1]:

$$\hat{X} = D \times \hat{\psi}$$

۴- یافتن درایه های غیر صفر ماتریس $\hat{\psi}$.

۵- برای هر درایه‌ی غیر صفر $\varphi(i,j)$، ۴ بیت کم ارزش قسمت اعشاری $\varphi(i,j)$ را یافته و بر اساس غلبه تعداد بیت صفر یا یک، بیت k ام پیام ($P'_k$) را بازیابی می کنیم.

---

شکل۲: الگوریتم بازیابی بیت های پیام مخفی

در این مرحله انتخاب سایز بلوک‌ها و تعداد ضرایب غیر صفر بسط تنک داده‌ها، در مرحله تخمین دیکشنری به روش آزمون و خطا به نحوی انتخاب می‌شود که ظرفیت مورد نظر برای درج پیام مخفی حاصل شود. الگوریتم پیشنهادی را برای درج پیام‌های شبه تصادفی با سایز ۳ کیلو بیت به کار می‌بریم. در مرحله تخمین دیکشنری توسط الگوریتم KSVD تعداد اتم‌های دیکشنری و تعداد ضرایب غیر صفر بسط تنک به ترتیب ۱۲۹ و ۳۱ انتخاب شدند. در نتیجه ظرفیت درج ۳/۹۷ کیلو بایت است. شکل ۳ نشان دهنده تصویر پوشش و

# نهان‌نگاری بر اساس تجزیه تنک بلوک‌های تصویر

تصویر نهان نگاری شده و هیستوگرام این تصاویر است. نسبت سیگنال به نویز تصویر نهان‌نگاری‌شده dB۳۰ است که نشانگر شفافیت بالای الگوریتم است.

روش پیشنهادی را می‌توان با روش های جایگزینی بیت کم ارزش و روش های بر مبنای تجزیه SVD مقایسه کرد. مزیت این روش نسبت به روش جایگزینی بیت کم ارزش عدم تغییر هیستوگرام و در نتیجه عدم تشخیص وجود پیام مخفی توسط حمله کننده است. مزیت روش پیشنهادی نسبت به روش های مبتنی بر تجزیه SVD ظرفیت بیشتر این روش است. زیرا در روش های مبتنی بر SVD درج را در برخی از مقادیر تکین انجام می دهند که حداکثر ظرفیت (چنانچه در تمامی مقادیر تکین درج انجام شود) برابر $J \times n$ خواهد بود که J تعداد بلوک‌ها و n سایز بلوک ها است.

تصویر نهان نگاری شده شفافیت مطلوبی دارد ولی در روش‌های مبتنی بر تجزیه SVD که درج در مقادیر تکین غیر صفر انجام می‌شود، حد بالای ظرفیت نیز به ندرت قابل دسترس است.

چنانچه به تصویر نهان‌نگاری شده نویز فلفل نمکی با نرخ سیگنال به نویز dB۱۳ بیفزاییم نرخ خطای بازیابی پیام مخفی حدود ۹٪ افزوده می شود. شکل ۴ تصویر نهان‌نگاری‌شده با افزودن نویز فلفل نمکی را نشان می‌دهد. عیب الگوریتم نهان نگاری مطرح شده در این مقاله این است که نرخ خطای بازیابی پیام حدود ۳۹٪ است. علت خطای بازیابی روش مذکور این است که پیام مخفی در ۴ بیتِ قسمت اعشاری ضرایب بسط تنک بلوک های تصویر درج می‌شود و بازیابی بیت های مخفی بستگی به دقت یافتن ضرایب بسط تنک توسط الگوریتم تجزیه تنک مورد استفاده دارد.

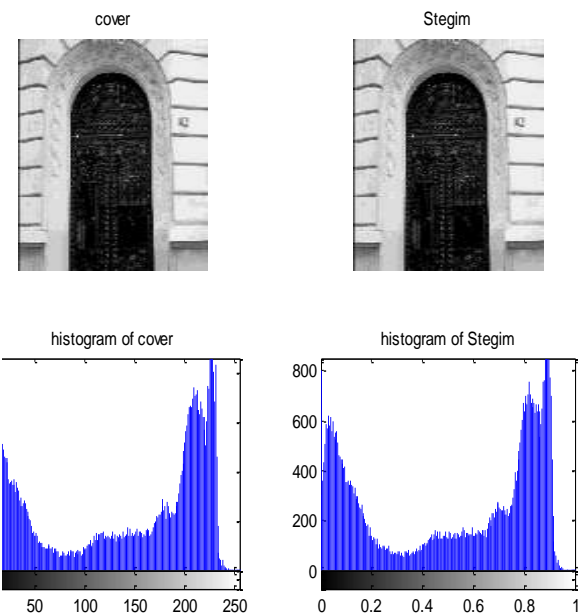

شکل۳:سطر اول: تصاویر پوشش و نهان نگاری شده. سطر دوم: هیستوگرام تصاویر پوشش و پنهان نگاری شده.

همانطور که گفته شد حد بالای ظرفیت روش پیشنهاد شده در این مقاله $J \times \left(\dfrac{n^2}{2}\right)$ است. حد بالای ظرفیت در روش پیشنهادی ما به راحتی قابل دسترس است و با درج در تمامی ضرایب نیز

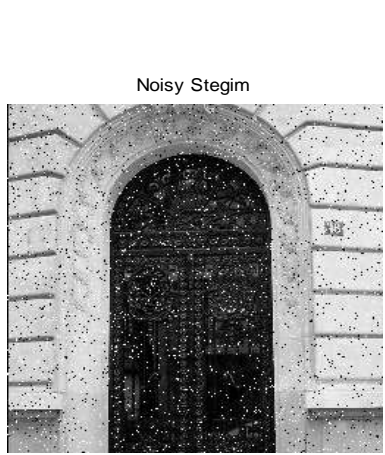

شکل۴: تصویر نهان نگاری شده با افزودن نویز فلفل نمکی

(SNR=13 dB[1])

برای کاهش نرخ خطای بازیابی می‌توان با استفاده از روش‌های کدینگ متفاوت، از بیت‌های تصحیح خطا استفاده کرد. هزینه‌ای که برای تصحیح خطا به این روش می‌پردازیم کاهش ظرفیت است که بسته به نوع کدینگ مورد استفاده برای کاهش خطا، ظرفیت درج پیام نیز متفاوت خواهد بود. روش دیگر کاهش خطا استفاده از این

---
[1] Signal to Noise Ratio

روش در حوزه‌ی تبدیل به طور مثال حوزه تبدیل موجک و یا DCT است. در این حالت درج در بیت‌های پرارزش‌تر انجام می‌گیرد و در نتیجه اثر دقت یافتن ضرایب بسط تنک در بازیابی پیام مخفی کاهش می‌یابد و نرخ خطای بازیابی کم می‌شود.

## ۵- نتیجه گیری و پیشنهادات

در این مقاله یک روش نهان نگاری ارائه شده است که ایده‌ی استفاده از تجزیه تنک بلوک‌های تصویر بر روی دیکشنری تخمین‌زده‌شده از تصویر را ارائه می‌دهد. این روش شفافیت مطلوبی دارد و هیستوگرام تصویر تغییرات مشابه با روش‌هایی مانند درج در کم‌ارزش‌ترین بیت را ندارد و در مقابل حمله کشف وجود پیام مقاوم است. علاوه بر این، الگوریتم نهان‌نگاری پیشنهادشده در این مقاله بر روی سایر انواع سیگنال‌های پوشش مانند صوت نیز قابل انجام است.

به منظور بهبود نرخ خطای بازیابی پیام مخفی می‌توان از کدینگ تصحیح خطا استفاده کرد. در ادامه این کار می‌توان الگوریتم پیشنهادی را در حوزه تبدیل به کار برد. در این حالت می‌توان درج را در بیت‌های پرارزش‌تر انجام داد به نحوی که شفافیت نیز حفظ شود و در نتیجه اثر دقت یافتن ضرایب بسط تنک در بازیابی پیام مخفی کاهش می‌یابد و نرخ خطای بازیابی کم می‌شود. از طرفی به طور مثال می‌توان توسط این الگوریتم درج داده را در سطوح مختلف از تبدیل موجک انجام داد و بدین ترتیب ظرفیت را افزایش داد.

## ۶- مراجع

# یک روش برای بهبود سیستماتیک نهان‌نگاری $LSB^+$

کاظم غضنفری

دانشکده مهندسی کامپیوتر – دانشگاه صنعتی امیرکبیر

دانشکده مهندسی برق – دانشگاه صنعتی شریف

kazemmit@aut.ac.ir

## چکیده

هدف از نهان‌نگاری[1] ایجاد یک کانال محرمانه و امن بین فرستنده و گیرنده می‌باشد. روش‌های گوناگونی برای ایجاد این کانال موجود است. از طرف دیگر روش‌های زیادی برای تشخیص این کانال محرمانه ارائه شده است. یکی از رایج‌ترین و ساده‌ترین روش‌های ایجاد این کانال، درج اطلاعات محرمانه در بیت کم‌ارزش[2] مقادیر پیکسل‌های تصویر می‌باشد. این روش به دلیل اینکه مشخصات آماری و ادراکی تصویر را به مقدار زیادی تغییر می‌دهد، مورد توجه حملات گوناگونی قرار گرفته است. یکی از این حملات، حمله هیستوگرام[3] می‌باشد. برای مقابله با این حمله اخیراً روشی تحت عنوان $LSB^+$ ارائه شده است که از طریق درج عمدی مقداری اطلاعات اقدام به حفظ هیستوگرام تصویر می‌کند. از آنجایی که درج عمدی منجر به ایجاد خرابی‌های ادراکی و خرابی‌های مراتب بالاتر می‌شود و زمینه را برای حملات گوناگون فراهم می‌آورد، در این مقاله روش جدیدی بر مبنای روش $LSB^+$ ارائه می‌شود که منجر به ایجاد کمترین خرابی می‌شود. برای این منظور کلید جدیدی معرفی می‌شود که بر مبنای آن، از طریق قفل کردن تعدادی از پیکسل‌های تصویر سعی می‌شود هیچ گونه درج عمدی رخ نداده و کمترین خرابی‌های ادراکی به وجود آید. نتایج آزمایش‌ها نشان می‌دهد که روش ما بدون کاهش ظرفیت، توانسته است تا حد زیادی میزان خرابی‌های ناشی از درج را کاهش دهد. همچنین نتایج بدست آمده بیانگر این مطلب است که اگر هیستوگرام تصویر دارای نوسانات زیاد باشد، روش پیشنهادی نسبت به روش $LSB^+$ منجر به نتایج بهتری می‌شود.

واژه‌های کلیدی: ۱- نهان‌نگاری، ۲- نهان‌کاوی[4]، ۳- حمله هیستوگرام، ۴- درج در بیت کم ارزش

## ۱- پیشگفتار

با توجه به رشد سریع ارتباطات و اینترنت، انتقال داده‌ها در قالب دیجیتال امری مرسوم شده است. داده‌های دیجیتالی می‌توانند به شکل متن، تصویر، صدا و یا فیلم باشند. یک مسئله مهم در رابطه با انتقال اطلاعات دیجیتال از طریق اینترنت آن است که امنیت انتقال اطلاعات تضمین شود. بطور کلی هدف در پنهان‌نگاری، ارسال پیام‌های محرمانه بصورت مخفیانه روی شبکه می‌باشد. پیام‌های محرمانه می‌تواند ابتدا رمز شده، سپس در یک پوشش[5] مانند تصویر پنهان شده و روی شبکه ارسال شوند. منظور از پیام رمزشده، دنباله‌ای از کدهای فاقد معنی می‌باشد که مهاجم قصد دارد آن پیام را رمزگشایی کند. اگرچه این فرد ممکن است به دلیل نداشتن رمزگشا، قادر به رمزگشایی این پیام‌های فاقد معنی نباشد، ولی در صورت مشکوک شدن به وجود اطلاعات در پوشش، می‌تواند موجب خرابی داده‌ها یا قطع شدن فرآیند انتقال شود. در مقابل،

---

[1] Steganogaphy
[2] LSB method
[3] Histogram attack
[4] Steganalysis
[5] Cover

# یک روش برای بهبود سیستماتیک نهان‌نگاری $^+$LSB

هدف از نهان‌نگاری آن است که مهاجم متوجه وجود داده‌های مخفی و در پوشش نشود [1،2،3].

روش‌های گوناگونی برای پنهان‌نگاری اطلاعات در تصویر ارائه شده است. در یک دسته‌بندی، روش‌های نهان‌نگاری را بر اساس حوزه درج به دو دسته اصلی تقسیم می‌کنند: نهان‌نگاری در حوزه مکان[1] و نهان‌نگاری در حوزه تبدیل[2]. روش‌های نهان‌نگاری در حوزه مکان، اطلاعات محرمانه را به طور مستقیم در مقدار پیکسل‌ها درج می‌کنند درحالیکه در روش‌های نهان‌نگاری در حوزه تبدیل، ابتدا تصویر به کمک یکی از تبدیلات مانند $DCT$، $DWT$ و $DFT$ به حوزه فرکانس انتقال داده می‌شود، سپس عمل نهان‌نگاری در این حوزه انجام می‌شود. درج اطلاعات در تصویر منجر به خرابی‌های گوناگونی می‌شود و به طور کلی به سه دسته تقسیم می‌شوند: خرابی‌های ادراکی، آماری و ساختاری، که این خرابی‌ها زمینه تشخیص وجود پیام محرمانه در تصویر را برای مهاجمان فراهم می‌نمایند [1،2،3]. حمله هیستوگرام یکی از رایج‌ترین حملات می‌باشد که از خرابی‌های آماری به وجود آمده بر اثر درج اطلاعات در تصویر برای تحلیل استفاده می‌کند. بر اساس این که عمل نهان‌نگاری در چه حوزه‌ای انجام شده است، حمله هیستوگرام قابل انجام می‌باشد و ما نیازمند روش‌هایی هستیم که هیستوگرام تصویر (یا ضرایب تبدیل) را پس از درج در حوزه مکان (یا تبدیل) حفظ کنند [5،7].

## 2- خرابی هیستوگرام

فرض کنید فقط بیت کم ارزش پیکسل‌ها برای درج اطلاعات محرمانه استفاده کنیم، اگر معادل دسیمال 7 بیت با ارزش یک پیکسل برابر $2i (0<i<127)$ باشد، در اینصورت اگر بیت کم ارزش صفر (یک) باشد مقدار پیکسل برابر $2i$ ($2i+1$) خواهد بود. حال اگر احتمال تغییر بیت کم ارزش (از صفر به یک و یا برعکس) را بر اثر درج پیام $P$ در نظر بگیریم، در اینصورت ماشین حالت محدود[3] شکل شماره 1 به وجود خواهد آمد. بر اساس این ماشین حالت، اگر $h_{2i}$ و $h_{2i+1}$ به ترتیب فراوانی پیکسل‌های با مقدار $2i$ و $2i+1$ باشند، در اینصورت انجام عمل درج در بیت کم ارزش تصویر باعث می‌شود که فراوانی پیکسل‌های با مقدار $2i$ و $2i+1$ به ترتیب به $h^*_{2i}$ و $h^*_{2i+1}$ تغییر پیدا کند که از روابط زیر بدست خواهند آمد:

$$h^*_{2i} = (P)h_{2i+1} + (1-P)h_{2i}$$

$$h^*_{2i+1} = (P)h_{2i} + (1-P)h_{2i+1}$$

$$|h^*_{2i} - h^*_{2i+1}| = |1-2P| \, |h_{2i} - h_{2i+1}|$$

با توجه به اینکه پیام قبل از درج رمز می‌شود و پیام رمز شده به صورت شبه‌تصادفی بوده و فراوانی رخ دادن صفر و یک برابر می‌باشد، به طور متوسط نیمی از بیت‌های کم‌ارزش تغییر می‌کنند. به عبارت دیگر $|1-2P|$ عددی مثبت و کوچک خواهد شد که در نهایت باعث می‌شود که اختلافِ فراوانیِ دو مقدار $2i$ و $2i+1$ کاهش پیدا کند [7].

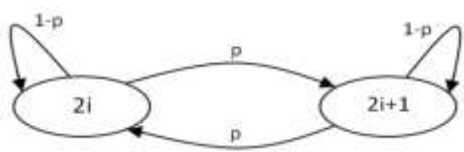

**شکل 1- ماشین حالت به وجود آمده از تغییر بیت کم‌ارزش.**

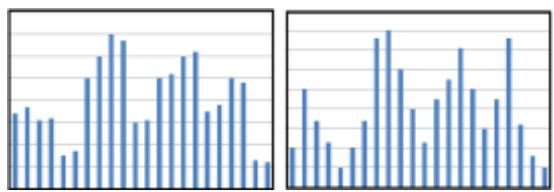

**شکل 2- هیستوگرام تصویر قبل از درج (سمت راست) و بعد از درج (سمت چپ).**

جهت توضیح بیشتر شکل شماره 2 را در نظر بگیرید، در این صورت اگر هیستوگرام تصویر قبل از درج اطلاعات محرمانه به صورت شکل شماره 2 (سمت راست) باشد، هیستوگرام تصویر بعد از درج به صورت شکل شماره 2 (سمت چپ) خواهد شد.

## 3- روش‌های مقابله با خرابی هیستوگرام

روش‌های گوناگونی برای حفظ و بازسازی هیستوگرام ارائه شده‌اند که هر کدام دارای مزایا و معایبی می‌باشند. در روش [8] بر اساس

---

[1] Special Domain
[2] Transform Domain
[3] Finite state machine

این که از چند بیت کم ارزش برای درج استفاده خواهد شد، تعدادی گروه تشکیل می‌شود، که در هر گروه bin هایی در یک واحد[1] قرار می‌گیرند. به طور مثال در صورت استفاده از یک بیت برای عمل درج، تعداد ۱۲۸ گروه و اگر از ۲ بیت برای درج استفاده شود، تعداد ۶۴ گروه به وجود خواهد آمد. بعد از مشخص‌شدن این گروه‌ها، توزیع هر یک از گروه‌ها به دست آمده و سپس پیام به قطعات کوچکی تقسیم می‌شود. بر اساس توزیعی که هر قطعه پیام دارد، گروه مناسب برای درج آن قطعه انتخاب شده و عمل درج در آن صورت می‌پذیرد. در این روش چون گروهی یافتن که توزیعش با توزیع قطعه پیام برابر باشد، بسیار مشکل و به ندرت اتفاق می‌افتد، هیستوگرام تصویر کاملاً حفظ نمی‌شود. همچنین چون برای درج هر قطعه باید گروهی پیدا شود که بیشترین تطابق را با توزیع قطعه پیام دارد، روش کندی می‌باشد. در روشی دیگر[2] [۹] به جای اینکه عمل درج بیت پیام به طور مستقیم در بیت کم‌ارزش انجام شود، اگر بیت پیام با بیت‌کم‌ارزش برابر باشد، تغییری اعمال نمی‌شود، ولی اگر برابر نباشد، بر اساس مقدار احتمالی که به صورت پارامتر مشخص می‌شود، بیت‌کم‌ارزش را یک واحد کم یا اضافه می‌کنند. مشخص است که با کاهش یا افزایش به اندازه یک مقدار پیکسل مربوطه، بیت‌کم‌ارزش برابر بیت پیام خواهد شد. این روش به جای اینکه سعی در حفظ هیستوگرام تصویر داشته باشد، بیشتر سعی دارد از نرم شدن آن ممانعت نماید، همچنین چون کاهش به مقدار یک ممکن است باعث شود دو بیت هر پیکسل تغییر یابند، زمینه را برای سایر حملات از جمله حملات آماری مراتب بالا[3] و حملات ساختاری و ادراکی فراهم می‌آورد. همچنین روشی در [۶] ارائه شده است که در آن، در صورتی که نیاز به تغییر بیت‌کم‌ارزش باشد، از طریق جابه‌جایی ضرایب در عوض تغییر آنها، هیستوگرام تصویر یا ضرایب تبدیل مربوطه حفظ می‌شوند. در روشی دیگر [۴] سعی می‌شود از طریق درج عمدی تعدادی صفر یا یک، هیستوگرام تصویر حفظ شود. در واقع در این روش از طریق قربانی‌کردن تعدادی پیکسل، هیستوگرام تصویر حفظ می‌شود. این روش به طور کامل هیستوگرام تصویر را حفظ می‌کند، ولی عیب عمده آن این است که درج عمدی باعث خرابی تصویر به لحاظ ادراکی شده و زمینه را

برای انواع حملات فراهم می‌نماید. در این مقاله سعی می‌شود با ارائه روشی جدید، تعداد پیکسل‌هایی که باید در فرآیند عمدی تغییر یابند کاهش داده شوند. برای این منظور با معرفی کلید جدیدی، قبل از انجام فرآیند درج، تعدادی از پیکسل‌ها را قفل می‌کنیم، به طوری که این پیکسل‌ها در فرآیند درج شرکت نمی‌کنند. جزئیات مربوط به این روش در بخش بعد شرح داده می‌شود.

## ۴- روش پیشنهادی

قبل از اینکه به بیان روش خودمان بپردازیم، به طور خلاصه روش $LSB^+$ را که مقدمه روش پیشنهادی می‌باشد بیان می‌کنیم. این روش هر سطح روشنایی در تصویر را یک bin در نظر می‌گیرد. هچنین بر اساس این که از چند بیت‌کم‌ارزش ($S$ عدد بیت) برای درج استفاده می‌شود، به تعریف "واحد" می‌پردازیم. در واقع هر واحد شامل تعدادی bin است که اگر $S$ بیت‌کم‌ارزش پیکسل‌ها را تغییر دهیم، مقدار آن پیکسل همچنان در یکی از bin های آن واحد باقی بماند. شکل شماره ۳ تعریف bin و واحد را برای حالتی که $S=1,2$ باشد نمایش می‌دهد.

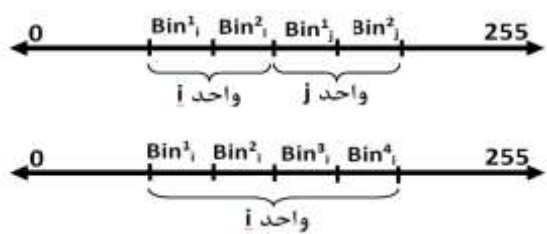

شکل ۳- تعریف واحد و bin برای $S=1$ (بالا) و $S=2$ (پایین).

نهان‌نگاری در فرستنده به این صورت انجام می‌شود که بعد از محاسبه هیستوگرام، طبق کلید درج، فرآیند درج پیام در هر واحد آنقدر ادامه می‌یابد تا تمام پیکسل‌های یکی از bin های آن واحد جهت درج پیام مورد استفاده قرار گیرد. بعد از این نمی‌توان پیامی را در این واحد درج نمود و باید سراغ بقیه واحدها برویم. از آنجایی که درج در واحد مربوطه باعث می‌شود که فراوانی bin های آن واحد تغییر کند، جهت بازسازی هیستوگرام از پیکسل‌های واحد مربوطه که برای درج استفاده نشده‌اند، استفاده کرده و با درج

---

[1] Unit
[2] LSB±
[3] High order statistics

# یک روش برای بهبود سیستماتیک نهان‌نگاری $LSB^+$

عمدی تعدادی صفر یا یک، اقدام به بازگرداندن هیستوگرام به شکل اولیه آن می‌کنیم. عمل استخراج در گیرنده به این صورت انجام می‌شود که بعد از بدست آوردن هیستوگرام، بر اساس کلید درج، استخراج اطلاعات از هر واحد آنقدر ادامه می‌یابد تا بیت‌کم‌ارزش تمام پیکسل‌های یکی از $bin$های آن واحد خوانده شود. بعد از آن مطمئن هستیم که هیچ پیام محرمانه‌ای در آن واحد وجود ندارد و لذا به سراغ واحد بعدی می‌رویم.

یکی از مشکلات عمده روش $LSB^+$ این است که حفظ هیستوگرام از طریق درج عمدی اطلاعات منجر به ایجاد یکسری خرابی‌های آماری مراتب بالا و ادراکی می‌شود. ما روشی را ارائه دهیم که نه تنها هیستوگرام تصویر را حفظ نماید، بلکه منجر به خرابی‌های آماری و ادراکی به مراتب کمتری شود و ظرفیت آن نسبت به روش $LSB^+$ کاهش چندانی پیدا نکند.

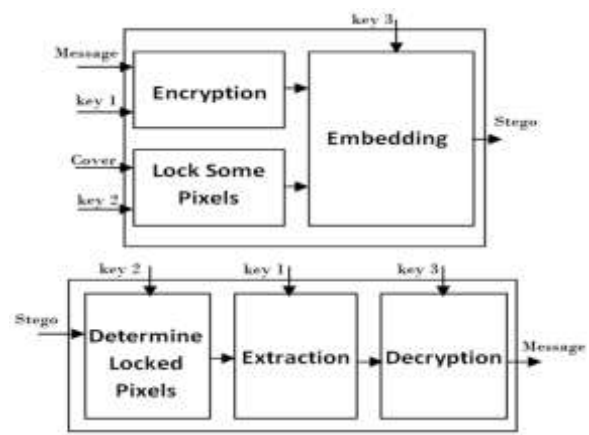

**شکل ۴- دیاگرام روش پیشنهادی برای فرآیند درج و استخراج پیام محرمانه.**

شکل شماره ۴ دیاگرام روش پیشنهادی را نمایش می‌دهد. در روش پیشنهادی ۳ عدد کلید استفاده شده است که در ادامه کاربرد هر یک از آنها را بیان می‌کنیم. ابتدا پیام محرمانه با استفاده از روش‌های رمزنگاری به کمک کلید اول رمز می‌شود. در رابطه زیر $Message^*$ پیام رمز شده می‌باشد:

$$Message^* = Encrypt(Message, key1)$$

همچنین قبل از انجام عمل نهان‌نگاری، هیستوگرام تصویر را بدست آورده و اختلاف فراوانی $bin$های هر واحد محاسبه می‌کنیم. سپس از طریق تعریف کلید دوم سعی می‌شود که تعدادی از پیکسل‌هایی که در این واحد قرار گرفته‌اند قفل شوند. بدون از دست دادن عمومیت، واحد $i$ را در نظر بگیرید. مشخص است که پیکسل‌های با مقدار $2i$ و $2i+1$ در این واحد قرار می‌گیرند. همچنین مجدداً بدون از دست دادن عمومیت فرض کنید که $h_{2i}>h_{2i+1}$ باشد. در این صورت اگر اختلاف فراوانی این دو $bin$ برابر $A_i$ باشد، سعی می‌شود به کمک کلید دوم تعداد $A_i$ پیکسل از تصویر که دارای مقدار $2i$ می‌باشند، قفل شوند:

$$Cover\hat{} = Lock(Cover, A_i, key2), \quad i=0,1,...,127$$

در رابطه فوق $Cover\hat{}$ تصویری است که در آن به ازای هر واحد $i$ تعداد $A_i$ پیکسل با مقدار $2i$ قفل شده‌اند و در صورتی که در مرحله بعد برای فرآیند درج انتخاب شوند (به کمک کلید سوم)، عمل نهان‌نگاری در آنها انجام نمی‌شود و انتخاب‌کننده پیکسل‌ها برای درج در صورت مواجه شدن با این پیکسل‌ها، بدون در نظر گرفتن آن‌ها سراغ انتخاب بعدی خود می‌رود. فرآیند قفل‌کردن باعث می‌شود که تعداد پیکسل‌های با مقدار $2i$ و $2i+1$ که برای درج‌کردن باقی‌مانده‌اند برابر شوند. به عبارت دیگر اگر $h\hat{}_{2i}$ و $h\hat{}_{2i+1}$ به ترتیب بیانگر فراوانی پیکسل‌های با مقدار $2i$ و $2i+1$ در تصویر $Cover\hat{}$ باشند که قفل نشده‌اند، داریم:

$$h\hat{}_{2i} - h\hat{}_{2i+1} = 0, \quad i=0,1,...,127$$

حال، عملیات درج در $bin$های هر واحد $i$ را بر اساس کلید سوم آنقدر ادامه می‌دهیم تا در یکی از $bin$های آن واحد $h\hat{}_{2i}$ عدد صفر یا یک درج شود:

$$Stego^+ = Embedding(Cover\hat{}, Message^*, key3)$$

چون پیام رمز شده به صورت شبه تصادفی بوده و احتمال رخدادن صفر و یک در آن برابر است و همچنین تعداد پیکسل‌های با مقدار $2i$ و $2i+1$ که بدون قفل می‌باشند برابرند، انجام عملیات نهان‌نگاری

به روش پیشنهادی باعث می‌شود اختلاف اولیه بین فراوانی bin های یک واحد تقریباً ثابت بماند و حمله هیستوگرام با شکست مواجه شود. در رابطه زیر L طول پیام رمز شده می‌باشد:

$$P(Message^*_i=1) \approx P(Message^*_i=0) \quad , \quad i=0,1,...,L$$

$$|h_i^+ - h_i| = \alpha_i \quad , \quad i=0,1,...,255$$

در رابطه فوق $h_i$ تعداد پیکسل های با مقدار i در تصویر Cover و $h_i^+$ تعداد پیکسل‌های با مقدار i در تصویر $Stego^+$ می‌باشد. لازم به ذکر است که با توجه به دلایل فوق، علی‌رغم این که هیستوگرام تصویر تقریباً تغییری نمی‌کند، ولی می‌توان برای حفظ کامل آن، از درج عمدی در پیکسل‌های بدون قفل باقیمانده در هر واحد استفاده نمود؛ در اینصورت :

$$Stego^* = intentional\_Embedding(Stego^+, \alpha_{2i}, \{0,1\})$$
$$, i=0,1,...,127$$

برای این منظور اگر رابطه $h_i^+ - h_i$ منفی (مثبت) شد، $\alpha_{2i}$ عدد صفر (یک) در پیکسل‌های بدون قفل که در مرحله درج استفاده نشده‌اند درج می‌کنیم. در اینصورت هیستوگرام تصویر کاملاً حفظ خواهد شد:

$$|h_i^+ - h_i| = 0 \quad , \quad i=0,1,...,255$$

عمل استخراج پیام می‌تواند به آسانی انجام شود. برای این منظور بعد از محاسبه هیستوگرام تصویر و اختلاف فراوانی bin ها، ($A_i$, $i=0,1,...,127$) پیکسل‌های قفل‌شده در هر واحد را بر اساس کلید دوم تعیین می‌کنیم:

$$Stego^\wedge = Lock(Stego^*, A_i, key2) \quad , \quad i=0,1,...,127$$

سپس به کمک کلید سوم اقدام به استخراج بیت‌های پیام می‌نماییم:

$$Message^* = Extraction(Stego^\wedge, key3)$$

توجه شود، همانند فرآیند درج، درصورتی که انتخاب‌کننده پیکسل‌ها برای استخراج با پیکسل‌های قفل‌شده مواجه شد، بدون در نظر گرفتن آن پیکسل سراغ انتخاب بعدی خود می‌رود. در نهایت عمل استخراج پیام را برای هر واحد را آنقدر ادامه می‌دهیم تا به تعداد $h_{2i}$ (یا $h_{2i+1}$) صفر (یا یک) استخراج کنیم. بعد از استخراج پیام، می‌توان پیام اصلی را به کمک رمزگشا بر اساس کلید اول بدست آورد:

$$Message = Decrypt(Message^*, key1)$$

با توجه به این که روش پیشنهادی منجر به تغییرات کمتری در مقدار پیکسل‌های تصویر می‌شود، به نظر می‌رسد که این روش نسبت به روش $LSB^+$ تاثیر مخرب کمتری بر اطلاعات آماری مرتبه دوم داشته باشد. اطلاعات آماری مرتبه دو، همبستگی پیکسل‌ها را دوبه دو در نظر می‌گیرد. یکی از بهترین روش‌های محاسبه این همبستگی استفاده از ماتریس‌های هم وقوعی یا co-occurrence می‌باشد. ماتریس هم وقوعی یک ماتریس دوبعدی است که مدخل‌های آن بیانگر میزان همبستگی سطوح خاکستری هر دو پیکسل با فاصله و جهت ثابت است. زوج $\Delta_x \Delta_y$ فاصله و جهت را توصیف می‌کند. در این صورت برای تصویر g، مدخل $c_{i,j}$ ماتریس هم‌وقوعی برابر تعداد حالاتی است که در آن‌ها به ازای $\Delta_x \Delta_y$ رابطه زیر برقرار باشد:

$$[g(x,y)=i] \quad and \quad [g(x+\Delta_x, y+\Delta_y)=j]$$

ماتریس فوق به ازای هر زوج $\Delta_x \Delta_y$ محاسبه می‌شود. از آنجایی که بدست آوردن ماتریس‌های فوق برای هر $\Delta_x \Delta_y$ خیلی هزینه‌بر است، معمولاً $\Delta_x \Delta_y$ طوری تعیین می‌شوند که همبستگی پیکسل‌های مجاور در نظر گرفته شوند. بر اساس مطلب فوق، واضح است هر چه پیکسل‌های بیشتری دچار تغییر شوند، اطلاعات آماری مرتبه دوم بیشتر تحت تاثیر قرار می‌گیرد. بنابراین به نظر می‌رسد که روش پیشنهادی تغییر کمتری در اطلاعات آماری مرتبه دوم را باعث شود.

## ۵- آزمایش‌ها

در آزمایش اول سعی داریم نشان دهیم که بین دو روش $LSB^+$ و روش پیشنهادی تفاوتی از نظر ظرفیت وجود ندارد. برای این منظور پیام با طول ۶۰ هزار بیت را در تعدادی از تصاویر به ابعاد ۲۵۰×۲۵۰ درج کردیم که نتایج آن در جدول ۱ مشاهده می‌شود.

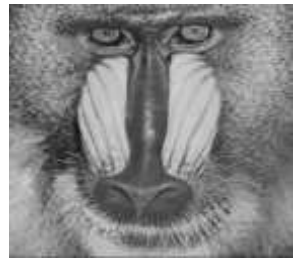
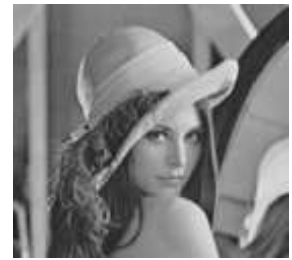

تصویر *Baboo*      تصویر *Lena*

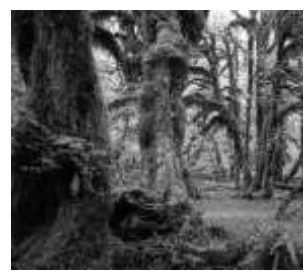
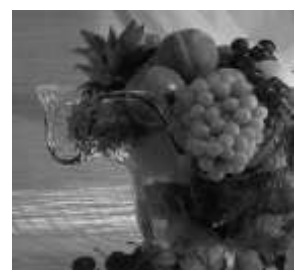

تصویر *Forest*      تصویر *Fruits*

**شکل ۵- مثال‌هائی از تصاویر مورد استفاده در آزمایش.**

همانطور که از جدول شماره ۱ مشخص است، هر دو روش تقریباً دارای ظرفیت یکسانی می‌باشند.

در آزمایش دوم، روش پیشنهادی را با روش $LSB^+$ از نظر میزان خرابی ادراکی حاصل از درج عمدی مورد مقایسه قرار می‌دهیم. برای این منظور بر حسب طول پیام درج شده در تصویر *Lena* با ابعاد ۵۰۰×۵۰۰، میزان خرابی حاصل را بدست آورده‌ایم. معیار مورد استفاده برای این آزمایش *PSNR* می‌باشد. این آزمایش منجر به منحنی شکل شماره ۶ شد. همانطور که این منحنی نشان می‌دهد، روش پیشنهادی منجر به خرابی کمتری، به خصوص در مورد پیام با اندازه بزرگ شده است.

**جـدول ۱- مقایسـه طـول پیـام قابـل درج در تصویر در روش پیشنهادی و روش LSB+.**

| تصــویر مــورد | حداکثر طول پیام در روش $LSB^+$ | حداکثر طول پیام در روش پیشنهادی |
|---|---|---|
| *Lena* | ۵۶۳۱۶ | ۵۶۲۹۱ |
| *Baboo* | ۵۳۱۶۴ | ۵۲۶۲۹ |
| *Fruit* | ۵۶۸۳۶ | ۵۶۶۲۸ |
| *Forest* | ۵۸۸۳۷ | ۵۵۶۹۴ |

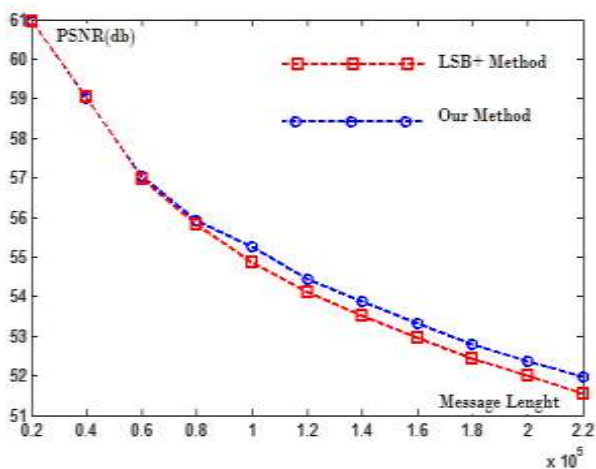

**شکل ۶- خرابی به وجود آمده در اثر درج به روش *LSB+* و روش پیشنهادی.**

هدف از آزمایش سوم آن است که نشان دهیم روش پیشنهادی برای تصاویری که هیستوگرام آنها دارای نوسانات زیاد است، نتـایج بهتری دارد و نسبت به روش $LSB^+$ منجر به تغییرات کمتری می‌شود. این آزمایش بر روی تعدادی تصویر که نوسانات هیستوگرام آن‌ها متفاوت بوده است انجام و به منحنی شکل ۷ منجر شد. برای بدست آوردن میزان نوسانات هیستوگرام از رابطه زیر استفاده شـده است:

$$Hist\_change = \sum_{i=0}^{127} |h_{2i+1} - h_{2i}|$$

همانطور که در شکل شماره ۷ دیده می‌شود، نه تنها میزان خرابی ناشی از درج به روش پیشنهادی کمتر از این اختلال در روش $LSB^+$ می‌باشد، بلکه میزان این خرابی در روش پیشنهادی تقریباً مستقل از نوسانات هیستوگرام است. امین در حالی است که روش $LSB^+$ با افزایش نوسانات هیستوگرام، خرابی‌های بیشتری را به وجود می‌آورد.

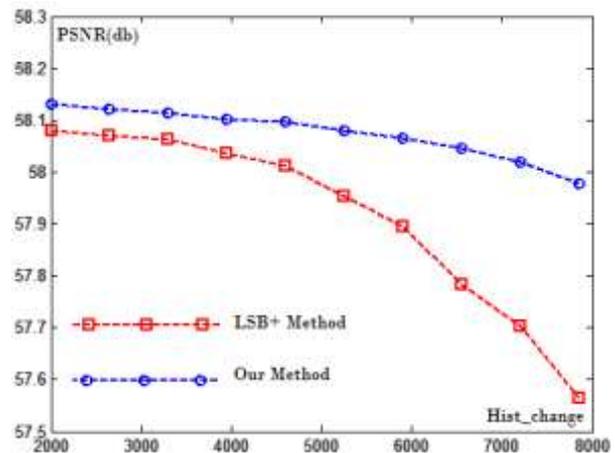

شکل ۷- تاثیر میزان نوسانات هیستوگرام تصویر بر خرابی به وجود آمده در اثر درج به روش $LSB^+$ و روش پیشنهادی.

## ۶- نتیجه گیری

درج اطلاعات محرمانه در بیت کم‌ارزش مقدار پیکسل‌های تصویر باعث تخریب اطلاعات آماری تصویر شده و زمینه را برای نهان‌کاوی موفق فراهم می‌آورد.. یکی از این موارد، خرابی هیستوگرام می‌باشد که زمینه را برای حمله رایج هیستوگرام فراهم می‌نماید. روش‌های زیادی برای حفظ و بازسازی هیستوگرام تصویر ارائه شده‌اند. در این مقاله روش جدیدی بر مبنای روش $LSB^+$ برای حفظ هیستوگرام ارائه شده است. برای این منظور، با تعریف کلید جدیدی، اقدام به قفل‌کردن تعدادی از پیکسل‌های تصویر کردیم. عملیات درج باید در پیکسل‌هایی انجام شود که در مرحله قبل قفل نشده‌اند. بعد از انجام عمل درج اطلاعات محرمانه، از طریق درج عمدی تعدادی صفر یا یک (که تعداد آنها در روش پیشنهادی به مراتب کمتر از این تعداد در روش $LSB^+$ است) سعی در بازسازی کامل هیستوگرام داشتیم. نتایج آزمایش‌ها بیانگر این است که روش پیشنهادی بدون از دست دادن ظرفیت، نسبت به روش $LSB^+$، منجر به خرابی کمی در تصویر می‌شود. همچنین نتایج آزمایش نشان می‌دهد که روش پیشنهادی برای تصاویری که هیستوگرام آن‌ها دارای نوسانات بیشتر است، به طور نسبی بهتر عمل می‌کند.

## ۷- مراجع

# ICA و کاربرد آن در نهان‌نگاری تصاویر دیجیتال

علیرضا شاه‌حسینی[1]، سید علی‌اصغر بهشتی شیرازی[2]

دانشگاه علم و صنعت ایران- گروه مخابرات امن

[1] ashahhoseini@ee.iust.ac.ir

[2] abeheshti@iust.ac.ir

## چکیده

آنالیز اجزاء مستقل (ICA) یک روش جداسازی کور منابع می‌باشد که در آن با استفاده از روشهای محاسباتی و آماری، یک سیگنال به مؤلفه‌ها یا اجزای سازنده‌اش (که برای ما مجهول می‌باشد) تجزیه می‌شود. فرضی که بر روی این مؤلفه‌های مجهول وجود دارد، غیر گاوسی بودن و استقلال آنها می‌باشد. سیگنال مورد تحلیل می‌تواند صوت، تصویر و یا فرضا یک سیگنال مخابراتی باشد. اخیرا تکنیکهای ICA، کاربرد خوبی در عرصه نهان‌نگاری تصاویر دیجیتال پیدا کرده است و از آن به عنوان حوزه جاسازی واترمارک، ابزار بازیابی واترمارک و ابزار حمله به الگوریتم‌های نهان‌نگاری استفاده شده است. در این مقاله، به بررسی و تشریح این کاربردهای سه‌گانه و چگونگی استفاده از ICA در آنها پرداخته شده است.

## واژه های کلیدی

آنالیز اجزاء مستقل، نهان‌نگاری، جداسازی کور منابع، جاسازی واترمارک، بازیابی واترمارک، حمله

## ۱- مقدمه

یکی از دغدغه‌های بشر ، حفظ حقوق معنوی آثار و اطمینان از تقلبی و ساختگی‌نبودن اسناد می‌باشد. با مطرح‌شدن رسانه دیجیتال ، اهمیت این مبحث بیش از پیش گردید چرا که کپی‌برداری رسانه دیجیتال ( اعم از عکس ، موسیقی و امثالهم ) بسیار راحت‌تر از قبل بوده و امکان جعل و تغییر آن نیز راحت‌تر از گذشته می‌باشد. قراردادن اطلاعات اثبات‌کننده مالکیت یک اثر دیجیتال ( که از آن تعبیر به "پیام" یا "واترمارک" می‌شود ) در داخل آن اثر ( که از آن به عنوان سیگنال حامل یا Cover تعبیر می‌شود ) ، واترمارکینگ دیجیتال نامیده می شود. واترمارک می‌تواند یک رشته تصادفی یا یک تصویر باینری یا Logo باشد [۱] . این جاسازی اطلاعات می تواند بصورت پیدا[1] یا ناپیدا[2] باشد [۲] . حالتی که از واترمارک غیر قابل رؤیت استفاده می شود، مرسوم‌تر می‌باشد و از آن تعبیر به " نهان نگاری" می شود. نهان نگاری، مفهومی متفاوت از استگانوگرافی[3] می باشد. اگرچه استگانوگرافی نیز یک روش جاسازی پیام مخفی در سیگنال حامل می‌باشد، اما بایدهای آن متفاوت می باشد. امروزه، نهان‌نگاری و استیگانوگرافی به یکی از مباحث مطالعاتی روز تبدیل شده است. تعریف پروژه‌های بین‌المللی مانند CERTMARK [۳] و ECRYPT [۴] گواه این مطلب می‌باشد .

در حالت کلی ، یک سیستم نهان نگاری را می‌توان معادل یک سیستم مخابراتی دانست. در حقیقت، همانطور که در شکل ۱ نیز نشان داده شده است، یک سیستم نهان‌نگاری را می‌توان یک سیستم مخابراتی دارای اطلاعات جانبی دانست [۵] که در آن

---

[1] Visible
[2] Imperceptible
[3] Steganography

# ICA و کاربرد آن در نهان نگاری تصاویر دیجیتال

سـیگنال حامـل، نقشـی مشـابه کانـال مخـابراتی، بلـوک جاسـازی واترمارک، نقش فرسـتنده و بلـوک بازیابی واترمارک، نقش گیرنـده را برعهده دارد. حملات اعمال‌شده به سیگنال نهان‌نگاری شـده را نیز می‌توان معادل نویز در کانال مخابراتی دانست. بـا ایـن اوصـاف، کاری که بلوک جاسازی واترمارک باید انجـام دهـد ایـن اسـت کـه رشته بیت پیام را بگونه‌ای مقاوم در برابر نویزها و اعوجاجات کانـال، بر روی کانال ارسال نماید. متقابلا وظیفه بلوک بازیابی واترمارک نیز تعیین وجود یا بازیابی واترمارک از روی سیگنال دریافتی می‌باشد.

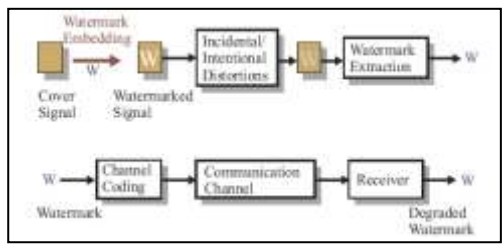

**شکل۱) تناظر بین الگوریتم نهان نگاری و سیستم مخابراتی [۵]**

مهاجم در یک سیستم نهان‌نگاری نیز نقش یک شـنودکننده کانـال مخابراتی و یا یک اعمال‌کننده نویز و اعوجاج به سیگنال مخابراتی را برعهده دارد.

در این مقاله، به بررسی و تشریح کاربردهای تبدیل $ICA^1$ در عرصه نهان‌نگاری تصاویر دیجیتال پرداخته شده است. ICA تبدیلـی مـی باشـد کـه بـردار مشــاهدات $X=(x_1,\ldots,x_m)^T$ را بـه بــردار $S=(s_1,\ldots,s_n)^T$ که در آن $s_i$ ها حداکثر استقلال آماری را نسبت به هم دارند، می‌نگارد. امروزه، تبدیل ICA به یکی از پرکـاربردترین و پرطرفدارترین ابزارهای پردازش سیگنال تبدیل شده است کـه ایـن امر به دلیل ویژگی‌های ساختاری این تبـدیل مـی‌باشـد. در عرصـه نهان‌نگاری، از تبدیل ICA به عنوان حوزه جاسازی واترمارک، ابـزار بازیابی واترمارک و ابزار حمله به الگوریتم‌های نهان‌نگاری، استفاده شده است که از این میان، کاربرد به عنوان ابـزار حملـه، از اهمیـت ویژه‌ای برخوردار می‌باشد. در ادامه، ابتدائاً به بررسـی تبـدیل ICA و مشخصات آن پرداخته شده است. تبدیل ICA را مـی‌تـوان در حالت کلی متشکل از دو جزء تقریبا مستقل از هم تابع هزینه[2] و الگوریتم بهینه‌سازی دانست. در ادامه این بخش به بررسی برخی از توابع هزینه و الگوریتم‌های بهینه‌سازی استفاده شده پرداخته شده است. بخش سوم مقاله به بررسـی Fast-ICA کـه در حـال حاضـر پرکاربردترین روش آنالیز اجزای مستقل مـی‌باشـد، اختصـاص داده

شده است. بررسـی اسـتفاده از تبـدیل ICA بـه عنـوان حـوزه جاسـازی واترمارک، بررسی قابلیت این تبدیل در بازیابی واترمـارک و بررسـی امکان استفاده از این تبدیل به عنوان ابزار حملـه بـه الگـوریتم‌هـای نهان‌نگاری موضوعاتی است که در بخشهای بعدی این مقالـه بـه آن پرداخته شده است.

## ۲- آنالیز اجزاء مستقل

یکی از مسـائل مطـرح در آمـار، پـردازش سـیگنال و شـبکه هـای عصبی، پیدا کردن حوزه تبدیل مناسب برای اطلاعات می باشـد. در پردازش الگـو[3]، فشرده‌سازی اطلاعات، حذف نویز و بسـیاری مـوارد دیگر ما به دنبال یافتن تبدیل‌های مناسب می‌باشـیم. یـک تبـدیل مناسب، تبدیلی می‌باشد که با استفاده از آن می‌توان آنالیز اطلاعات را به بیشترین مقدار ممکن ساده نمود. یـافتن تبـدیل مناسـب، بـه معنای یافتن تابع f می باشد که بـردار $X=(x_1,\ldots,x_m)^T$ ( بـردار $m$ بعـدی ورودی ) را بـه بـردار $n$ بعـدی $S=(s_1,\ldots,s_m)^T$ - کـه دارای خواص مشخصی است می‌نگارد.
در بسیاری از موارد، f یک تابع خطی فرض می‌شود. در اینحالت S یک تبدیل خطی از X ، بصـورت S=WX مـی‌باشـد کـه در آن W ماتریس تبدیل بوده و باید تعیین گـردد. اسـتفاده از تبـدیل خطـی علاوه بر آنکه تصور مسئله را ساده تر می‌نماید، پیچیدگی محاسباتی را نیز کاهش می‌دهد. در بسیاری از تبدیل‌ها ماننـد تبـدیل DWT ، DFT و DCT ، ماتریس W یک ماتریس مشخص و ثابت می‌باشد. در حقیقت در اینجا بردار ورودی بر روی بردارهـای پایـه مشخصـی تصویر می‌شود. در دسته دیگر تبـدیل‌ها، همـین مـاتریس W نیـز از روی بـردار ورودی بدسـت مـی‌آیـد. در حقیقـت در ایـن دسـته از تبدیل‌ها، بردارهای پایه نیز از روی بردار ورودی تعیـین مـی‌گـردد و متناسب با آن تغییر می‌کند. محاسبه ماتریس W در ایـن دسـته از تبدیل‌ها نیازمند اطلاع از مشخصات آماری بـردار ورودی و حتـی در برخی موارد، توزیع آماری مؤلفه‌های ایـن بـردار مـی‌باشـد. چنـین اطلاعاتی، با داشـتن صـرفا یـک بـردار ورودی ( بـردار مشـاهدات ) حاصل نمی‌شود و نیازمند T بردار مشاهدات می‌باشـیم. واضـح است که هرچه T بزرگتر باشد، به تخمین بهتری از W دسـت مـی‌یابیم. مزیت این دسته از تبدیل‌ها، بهینگی آنها مـی‌باشـد. $PCA^4$ و ICA از جمله این تبدیل‌ها می‌باشند. معیار بهینگی تعریف‌شده در این تبدیل‌ها متفاوت می‌باشد. از این رو هریک از ایـن تبـدیل‌ها بـرای کاربردهای خاصی مناسب می‌باشند.
آنچه که تحت عنوان تبـدیل ICA شـناخته مـی شـود، اولـین بـار

---

[1] Independent Component Analysis
[2] Objective Function

[3] Pattern Recognition
[4] Principal Component Analysis

# ICA و کاربرد آن در نهان نگاری تصاویر دیجیتال

در ۱۹۸۰ و توسط J.Herault، C.Jutten و B.Aus مطرح گردید [۶]. این بحث، در اواسط دهه ۹۰ و پس از چاپ مقاله J.Bell و T.J.Sejnowski بسیار مورد توجه قرار گرفت. در این مقاله از دیدگاه infomax به حل مسأله پرداخته شده است [۷]. از این زمان به بعد، روشهای متعدد و بسیار مختلفی برای حل مسأله پیشنهاد شده است. نقطه عطف این جریان، کار انجام شده توسط S.Amari [۸] و J.Cardoso [۹] می‌باشد. این دو بصورت مستقل از هم و از دو نقطه شروع متفاوت، نشان دادند که گرادیان طبیعی[1] روش بهینه‌سازی است که می‌تواند ما را در جهت بیشترین کاهش یک تابع خطای داده شده در فضای ماتریسهای مربع ناویژه برساند [۱۰]. ارائه الگوریتم FastICA توسط A.Hyvarinen و J.Karhunen در [۱۱] و راندمان بالای محاسباتی این الگوریتم، زمینه ساز استفاده بسیار زیاد ICA در حل بسیاری از مسائل دیگر گردید. امروزه ICA به عنوان یک روش مطرح در جداسازی منابع، استخراج ویژگی، حذف نویز، نهان‌نگاری، تصویربرداری مغز، مخابرات و ... مطرح می باشد.

بنا به تعریف، ICA تبدیلی می باشد که بردار مشاهدات $X=(x_1,...,x_m)^T$ را به بردار $S=(s_1,...,s_n)^T$ که در آن $s_i$ ها حداکثر استقلال آماری را نسبت به هم دارند، می‌نگارد. اگر فرض کنیم که m سنسور داریم و مقدار هریک از مؤلفه‌های بردار X، معادل مقدار اندازه گیری شده توسط یکی از این سنسورها می‌باشد، در اینصورت مؤلفه i ام بردار خروجی الگوریتم ICA ( بردار S )، مقدار ارسال شده توسط منبع مستقل i ام می‌باشد. در یک بیان ساده، ICA مجموعه‌ای از الگوریتم‌ها و تکنیک‌ها می‌باشد که مجموعه‌ای از سیگنالهای دریافتی را بصورت ترکیبی خطی از چند منبع مستقل و نامعلوم تولیدکننده سیگنال مدل می‌نماید. این منابع می‌توانند حقیقی یا مجازی باشند. فلذا ICA را می توان تبدیلی دانست که بردار مشاهدات را به عناصر سازنده‌اش تجزیه نموده و ساختار مخفی اطلاعات را کشف می‌نماید. در [۱۲] اثبات شده است که در صورت برقراری شرایط زیر امکان اعمال تبدیل ICA و بدست آوردن ساختار مخفی بردار مشاهدات یا تشخیص عناصر سازنده آن وجود دارد:

1. تمام بردارهای $s_i$ ، مستقل از هم باشند.
2. تمام بردارهای $s_i$ بجز حداکثر یکی از آنها توزیع غیرگاوسی داشته باشد.
3. ماتریس A ( معکوس ماتریس W ) با رتبه کامل باشد.

در صورتی که تعداد سنسورها (m) ، حداقل برابر با تعداد مؤلفه های مستقل (n) باشد ، برقراری شرط سوم ، تضمین شده است [۱۲]. در تحلیل ICA، بردار مؤلفه‌های مستقل حاصله یکتا نبوده و دارای ابهامات زیر می باشد [۱۳]:

1. در این مدل امکان تعیین واریانس ( انرژی ) مؤلفه های مسقل وجود ندارد. دلیل این امر اینست که در اینجا S و A هر دو نامعلوم می باشند و:

$$X = \sum_i a_i s_i = \sum_i (\frac{1}{\alpha_i} a_i)(\alpha_i s_i) \quad (1)$$

2. ابهام در علامت وجود دارد. به عبارت دیگر با تغییر علامت مؤلفه مستقل، تغییری در مدل ایجاد نمی شود. با قرار دادن $\alpha_i = 1$ در رابطه بالا ، می‌توان این نکته را اثبات نمود.

3. امکان تعیین درجه مؤلفه های مستقل ندارد. چرا که قرار دادن ماتریس تبدیل P و معکوس آن در مدل تغییری ایجاد نمی نماید . به عبارت دیگر داریم:

$$X = AS = AP^{-1}PS \quad (2)$$

به عبارت دیگر همانطور که S یک بردار مؤلفه های مستقل می باشد ، PS نیز یک بردار مؤلفه های مستقل می‌تواند باشد . منشاء این ابهام نیز نامعلوم بودن توامان A و S می باشد.

در حالت کلی، الگوریتم‌های ICA مبتنی بر بیشینه یا کمینه کردن یک تابع می‌باشند. این تابع که از آن تحت عنوان تابع Objective ، کنتراست یا هزینه یاد می‌شود، تابعی است که مشخصات آماری مدل ICA مانند مقاومت و استحکام را تعیین می‌نماید. دو نوع تابع هزینه وجود دارد. توابع هزینه چندگانه، توابعی هستند که با بهینه‌سازی آنها تمام مؤلفه های مستقل را می‌توان با هم و در یک زمان تعیین نمود. بر عکس ، توابع هزینه یگانه توابعی می باشند که بهینه‌سازی آنها منجر به یافتن یکی از مؤلفه های مستقل می‌گردد. بنابراین بجای اینکه کل مدل ICA به یکباره تخمین زده شود، در هر مرحله تحت شرایط خاصی برداری مانند $w_i$ تخمین زده می شود بگونه‌ای که $w_i^T X$ یکی از مؤلفه های مستقل $s_i$ را تولید نماید. برای بهینه‌سازی نیز می توان از هریک از الگوریتم‌های کلاسیک مانند الگوریتم گرادیان، الگوریتم نیوتون و امثالهم استفاده نمود. الگوریتم بهینه‌سازی، تعیین‌کننده مشخصات ساختاری الگوریتم ICA مانند سرعت همگرایی، حافظه مورد نیاز و استحکام عددی الگوریتم می‌باشد. در حالت کلی، این دو بخش را می‌توان تاحدی مستقل از یکدیگر دانست و با انتخاب تقریبا مستقل از هم یک الگوریتم بهینه سازی و یک تابع هزینه می توان به الگوریتم‌های ICA مختلف رسید و یا الگوریتم‌های جدیدی طراحی کرد.

---

[1] Natural Gradient

## 3- Fast ICA

در حال حاضر، Fast ICA پرکاربردترین الگوریتم آنالیز اجزای مستقل موجود می‌باشد. این امر، بدلیل راندمان بالای محاسباتی این الگوریتم می‌باشد.

در اینجا ابتدائاً یک سری عملیات پیش‌پردازش انجام می‌شود. انجام این عملیات پیش از اجرای الگوریتم ICA ، موجب ساده‌تر شدن روند اجرایی الگوریتم می‌گردد. متمرکز کردن و سفیدسازی از جمله این عملیات می‌باشند. متمرکز کردن بردار ورودی، به معنای تبدیل بردار مشاهدات به یک بردار با میانگین صفر می‌باشد. برای این منظور، میانگین بردار مشاهدات را از آن کسر می‌نماییم. میانگین بردار مشاهدات، از میانگین‌گیری روی بردارهای مشاهدات $x(1)$ , $x(2)$ , ... , $x(T)$ بدست می‌آید. سفیدسازی نیز تبدیلی خطی می‌باشد که بردار مشاهدات را به برداری ناهمبسته و با واریانس یکه می‌نگارد. از آنجا که ICA قادر به تعیین واریانس مؤلفه های مستقل نمی‌باشد، بدون از بین رفتن عمومیت مسأله می‌توان فرض نمود که بردار مؤلفه‌های مستقل نیز یک بردار سفید می‌باشد. در حالت کلی، یک ماتریس متعامد، از $n(n-1)/2$ درجه آزادی برخوردار است . این در حالی است که یک ماتریس غیر متعامد دارای $n^2$ درجه آزادی می‌باشد. نتیجتا درصورت سفید سازی بردار مشاهدات قبل از اجرای الگوریتم ICA ، تعداد پارامترهایی که باید توسط الگوریتم ICA تخمین زده شود، تقریبا نصف خواهد شد . از آنجایی که عملیات پیش‌پردازش سفیدسازی یک فرایند ساده و سرراست بوده و پیچیدگی محاسباتی خاصی ندارد، پیچیدگی محاسباتی کل سیستم به شدت کاهش می یابد.

در Fast-ICA از توابع هزینه‌ی مبتنی بر بیشینه‌سازی خاصیت غیرگاوسی مؤلفه ها استفاده شده است. در حالت کلی، بردار مشاهدات X و بردار مؤلفه های مستقل S، رابطه‌ای بفرم $X = AS$ دارند (مدل ICA). از طرف دیگر، ما دنبال مولفه‌های مستقل هستیم که می توان آنها را بصورت ترکیبی خطی از $x_i$ ها تخمین زد. به عبارت دیگر، بردار مؤلفه‌های مستقل تخمین زده شده ($\hat{S}$) و بردار مشاهدات X، رابطه‌ای بفرم $\hat{S} = W^T X$ دارند. با ترکیب این دو رابطه خواهیم داشت :

$$\hat{S} = W^T AS \qquad (3)$$

یعنی بردار مؤلفه‌های مستقل تخمین زده شده را می‌توان بصورت ترکیب خطی بردار مؤلفه‌های مستقل واقعی دانست . در بهترین حالت $W^T A = I$ خواهد بود. در قضیه حد مرکزی بحث می‌شود که مجموع چند متغیر تصادفی غیرگاوسی، گاوسی‌تر از هریک از این متغیرهای تصادفی می‌باشد. به عبارت دیگر توزیع احتمال آن به توزیع احتمال گاوسی نزدیکتر می‌باشد. بنابراین با فرض اینکه هیچیک از مؤلفه‌های مستقل $s_i$ دارای توزیع گاوسی نمی‌باشند ( یکی از اصول ICA ) غیرگاوسی‌ترین توزیع برای $\hat{s}_i$، زمانی حاصل می‌شود که $\hat{s}_i = s_i$ گردد. نتیجتا اگر ماتریس W بگونه‌ای انتخاب گردد که خاصیت غیرگاوسی بردار $\hat{S}$ حداکثر گردد، ما در واقع تخمینی از بردار عناصر مستقل S بدست آورده‌ایم. در Fast ICA از همین نکته استفاده شده است.

یکی از معیارهای سنجش غیرگاوسی بودن یک بردار ، تابع نگنتروپی می‌باشد . نگنتروپی کمیتی است که از مفهوم آنتروپی در تئوری اطلاعات مشتق شده است. آنتروپی کمیتی است گویای میزان اطلاعات حاصله از مشاهده یک متغیر تصادفی می‌باشد . هرچه یک متغیر تصادفی غیرقابل پیش‌بینی‌تر و غیرنظام‌مندتر باشد، آنتروپی آن بزرگتر است . در تئوری اطلاعات اثبات می‌شود که متغیر تصادفی گاوسی، بزرگترین آنتروپی را در بین متغیرهای تصادفی دارای واریانس مساوی دارا می‌باشد. با این پیشینه، نگنتروپی یک بردار تصادفی بصورت زیر تعریف می‌شود:

$$J(\bar{Y}) = H(\bar{Y}_{gauss}) - H(\bar{Y}) \qquad (4)$$

که در آن $\bar{Y}_{gauss}$ بردار تصادفی گاوسی دارای ماتریس کواریانس برابر با بردار Y می‌باشد . رابطه نگنتروپی با اطلاعات متقابل به فرم زیر می‌باشد :

$$I(y_1,...,y_n) = J(\bar{Y}) - \sum_i J(y_i) + \frac{1}{2} \log \frac{\prod C_{ii}^Y}{\det C^Y} \qquad (5)$$

که در آن $C^Y$ ماتریس کواریانس Y بوده و $C_{ii}^Y$ عناصر قطری این ماتریس می‌باشد . اگر $y_i$ ها ناهمبسته باشند، ترم سوم رابطه فوق برابر صفر خواهد بود. در اینصورت داریم:

$$I(y_1,...,y_n) = J(\bar{Y}) - \sum_i J(y_i) \qquad (6)$$

تغییرناپذیری تابع نگنتروپی بر اثر تبدیل خطی [12] سبب می‌گردد که پیدا کردن جهت ماکزیمم نگنتروپی یعنی جهتی که المانهای مجموع $J(y_i)$ بیشینه گردد، معادل پیدا‌کردن حالتی است که اطلاعات متقابل کمینه گردد .

محاسبه نگنتروپی کار مشکلی می‌باشد. از این رو در عمل آنرا با کومولان های مرتبه بالا تخمین می‌زنند. در [44] در حالت میانگین صفر و واریانس واحد، تقریب زیر برای این تابع ارائه شده است :

$$J(y) \approx \frac{1}{12} \kappa_3(y)^2 + \frac{1}{48} \kappa_4(y)^2 \qquad (7)$$

که در آن $\kappa_i(y)$ کومولان درجه i ام متغیر تصادفی y می‌باشد. در

[45] نشان داده شده است که تخمین مبتنـی بـر کومـولان تابع نگنتروپی تخمین غیردقیقی بوده و در بسیاری از مواقع، حساس به لایه‌های بیرونی می‌باشد. تخمین بهتر این تابع به فرم :

$$J(y) \approx k_1(E\{G_1(y)\})^2 + k_2(E\{G_2(y)\} - E\{G_2(v)\})^2 \quad (8)$$

می‌باشد که در آن $G_1$ و $G_2$ توابع با درجه بزرگتر از دو بوده و $k_1$ و $k_2$ دو ثابـت مثبـت و $v$ یـک متغیـر گاوسـی اسـتاندارد دارای میانگین صفر و واریانس واحد می باشد. تقریب قبلـی را مـی تـوان حالت خاصی از این تقریب بازای $G_1 = y^3$ و $G_2 = y^4$ دانست. فرم ساده شده رابطه 8، بصورت:

$$J(y) \approx c(E\{G(y)\} - E\{G(v)\})^2 \quad (9)$$

می‌باشد. مزیت تقریب فوق، اینست که در آن تنهـا از یـک تـابع بـا درجه بالاتر از دو استفاده شده است. c در این رابطه ، یک ثابت مـی باشد. در Fast ICA، از همین رابطه ساده‌شده استفاده می‌شود. قرار دادن $G(y) = y^4/4$ در این رابطـه، مـا را بـه تقریـب مبتنـی بـر کرتیزوس، یکی از پرکاربردترین تقریبهای ارائه شـده بـرای تـابع نگنتروپی، می رساند. به راحتی می‌تـوان دیـد کـه کرتیـزوس یـک متغیـر تصـادفی گاوسـی برابـر صـفر، کرتیـزوس توزیـع هـای SuperGaussian مانند لاپلاسین، مثبت و کرتیزوس توزیعهـای SubGaussian مانند یکنواخت، منفی می‌باشـد [12]. پیچیـدگی محاسباتی کم و همگرایی خوب، از جمله مزایای ایـن تقریـب مـی‌باشد. مشکل این تقریب نیز در حساسیت بالا به tail هـای توزیـع و حساسیت بسیار کم نسبت به سـاختار میـانی توزیـع مـی‌باشـد. در [14] نشـان داده شـده اسـت کـه اسـتفاده از توابـع بـه فـرم $G(y) = -\exp(-y^2/2)$، $G(y) = (\log\cosh a_1 y)/a_1 \quad 1 \le a_1 \le 2$ منجر به افزایش مقاومت و استحکام الگوریتم می‌گردد.

هدف ما در FastICA، محاسـبه مـاتریس $W = [w_1, w_2, ..., w_n]$ در رابطه $\hat{S} = W^T X$ می‌باشد. بردار $w_i$ در اینجا، بردارای است کـه تابع نگنتروپی را بازای متغیر ورودی $\hat{s}_i = w_i^T X$، بیشینه مـی‌نمایـد. این کار با اجرای تکراری الگوریتم سریع نقطه ثابت زیر انجـام مـی‌شود [13]:

$$w_i^+ = E\{XG'(w_i^T X)\} - E\{G''(w_i^T X)\}w_i \quad (10)$$
$$w_i = w_i^+ / \|w_i^+\|$$

'G و "G در این رابطه به ترتیـب مشـتق اول و مشـتق دوم یکـی از توابع G تعریف شده در رابطه 9 می‌باشـند. الگـوریتم بهینه‌سـازی استفاده شده در فوق، یک الگوریتم بهینه‌سازی مبتنـی بـر گرادیـان می‌باشد. خوبی این روش بهینه‌سازی، همگرایی سریع آن می‌باشد.

در حالت کلی، بردارهای $w_i$ را می‌توان بصورت یکی‌یکی و یـا مـوازی محاسبه نمود. در روش اول، مؤلفه های مستقل یکی پس از دیگری تخمین زده می شوند. ایـن درحـالی اسـت کـه در روش دوم، همـه مولفه‌های مستقل با هـم تخمـین زده مـی‌شـوند. در اینصـورت بـه راحتی می‌توان مشاهده نمود که رابطه تکرار در حالت اول بفرم:

$$w_i^+ = E\{XG'(w_i^T X)\} - E\{G''(w_i^T X)\}w_i$$
$$w_i^+ = w_i^+ - \sum_{j=1}^{i-1}(w_i^{+T} w_j)w_j \quad (11)$$
$$w_i = w_i^+ / \|w_i^+\|$$

و در حالـت دوم، یعنـی حـالتی کـه بردارهـای $w_i$ بصـورت مـوازی تخمین زده می‌شوند، بفرم:

$$\forall i: w_i = E\{XG'(w_i^T X)\} - E\{G''(w_i^T X)\}w_i \quad (12)$$
$$W = (WW^T)^{-1/2} W$$

در می‌آید.

## 4- حوزه جاسازی واترمارک

در حالت کلی، برای جاسازی واترمارک در الگوریتم‌های نهان‌نگاری می‌توان از هر یک از حوزه‌های مکان و یا تبدیل استفاده نمود. یکـی از معیارهای طبقه‌بندی الگوریتم‌های نهان‌نگاری و یکـی از عوامـل تعیین‌کننده میزان مقاومت الگوریتم نهان‌نگاری در برابر حمـلات، همین حوزه جاسازی واترمارک می‌باشد. در بخشهای قبل، بحث شد که حوزه‌های تبدیل خطی و غیرخطی متفـاوتی وجـود دارد کـه هریک خواص ویژه‌ای را برای ما به ارمغـان مـی‌آورد و کاربردهـای خاص خود را دارد. حوزه تبدیل مـورد بحـث در ایـن مقالـه، حـوزه تبدیل خطی ICA می‌باشد. در ایـن بخـش از مقالـه، دسـته‌ای از الگوریتم‌های نهان‌نگاری که از تبدیل خطی ICA بـه عنـوان حـوزه جاسازی واترمارک استفاده می‌نمایند، مورد بررسـی قرارگرفتـه اسـت. ورود ICA به عرصه نهان‌نگاری از همین منظر بوده است [15]. در اینجا با استفاده از تجزیه ICA، سیگنال بـه حـوزه تبـدیل بـرده شده و واترمارک در ایـن حـوزه اعمـال مـی شـود. بازیـابی مجـدد واترمارک ( الگوریتم‌های Readable ) و یا اطمینـان از وجـود آن ( الگوریتم‌های Detectable ) نیز بطور مشابه در حوزه تبـدیل انجـام می شود. همانطور که بیان شد، یک سیسـتم نهـان نگاری را می توان معادل یک سیسـتم مخـابراتی دانسـت. سیسـتم مخابراتی معادل الگوریتم‌های نهان نگاری حوزه ICA، در شکل 2 نشان داده شده است. در این شکل، I معرف تصویر اصلی، S معرف تصویر اصلی که به حوزه ICA رفته است، m مبیـن واترمـارک کـه باید جاسازی شود، n بیانگر تخریب ناشی از حمـلات اعمـالی شـده بر تصویر نهان شده، $\hat{S}$ مبین تبدیل ICA تصویر نهان‌نگاری‌شده و $\tilde{S}$

# ICA و کاربرد آن در نهان نگاری تصاویر دیجیتال

بیانگر تبدیل ICA تصویر نهان‌نگاری‌شده دریافتی ( که حملاتی بـر روی آن اجرا شده است ) می‌باشد. همانطور که در شکل دیده مـی- شود، ضرایب حوزه ICA تصویر، کانال مخابراتی انتقال واترمارک ( بلوک خط چین ) می‌باشند. در این کانـال، حملـات و تخریبهـای اعمال شده بر تصویر نهان نگاری شده نقـش نـویز جمـع شـونده را خواهنـد داشـت. در حالـت کلـی، تبـدیل ICA یـک تبـدیل دارای پیچیدگی محاسباتی بالا می‌باشد. هرچه طول بردار مشاهدات (بردار ورودی تبدیل) بزرگتر باشد، بار محاسباتی نیز بیشتر خواهد بود. در عوض هرچه طول بردار مشاهدات کـوچکتر باشـد، کیفیـت و دقـت محاسبات کمتر خواهد بـود. تجزیـه تصویر اصـلی بـه بلـوک هـای ناهمپوشان و اعمـال تبـدیل ICA بر این بلوکهـا و جاسـازی واترمـارک در این مقادیر حوزه تبدیل، راه حل مورد استفاده به جهـت کـاهش بار محاسباتی در اغلب الگوریتم‌های نهان‌نگاری می‌باشد.

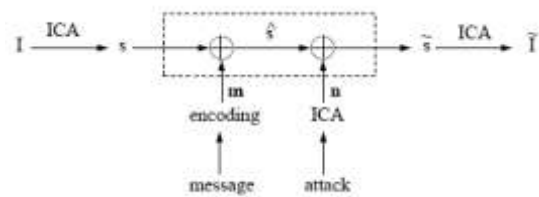

**شکل۲) سیستم مخابراتی معادل الگوریتم‌های نهان‌نگاری حوزه ICA [۱۶]**

اینکه بردارهای پایه حوزه ICA ، مشخصاتی مشابه سلولهای اولیه و ثانویه پوسته بینایی[1] در سیسـتم بینـایی انسـان دارنـد [۱۷] یعنـی localized بوده، میانگذر بوده و در فضای مکان متمرکز می‌باشند، یکی از علل اصلی توجه بـه ایـن حـوزه بـه عنـوان حـوزه جاسـازی واترمارک می‌باشد [۱۸]. چرا کـه مطـابقت بیشتر فرایند جاسـازی بـا HVS[2] ، نهان‌نگاری بهتر را به همراه می‌آورد. توجه به حوزه ICA به عنـوان حـوزه جاسـازی واترمـارک، پـس از ارائـه کـار Moulin و O'Sullivan در سال ۲۰۰۴ بود. آنها در مقاله خود بـا یـک تحلیـل تئوری اطلاعـاتی نشـان دادنـد کـه بیشـینه ظرفیـت نهـان‌نگـاری اطلاعات، تنها در حالی که کانالهای جاسازی دارای استقلال آماری باشند قابل دستیابی خواهد بود و هرگونه وابستگی آمـاری بـین ایـن کانالها سبب کاهش ظرفیت نهان‌نگاری خواهـد گشـت [۱۹]. بـر همـین اسـاس در [۱۶] اثبـات شـده اسـت کـه در یـک جاسـازی مشخص، در حالی که کانالهای جاسازی واترمارک دارای استقلال آماری باشند، شـاهد کمتـرین تخریـب ( بـر اثـر دسـته وسـیعی از حملات ) خواهیم بود.

الگوریتم F.Gonzalez [۱۵]، اولین و ساده‌ترین الگوریتم اسـتفاده- کننده از ICA در عرصه نهان‌نگاری مـی‌باشد. مشـابه روشهـای فشرده‌سازی مبتنی بر ICA، ایده اصلی در این مقاله امکان حذف و در نظر نگرفتن r مؤلفه ICA با انرژی کمتر سـیگنال مـی‌باشـد. بـا انتخاب مناسب r ، از این الگوریتم می‌توان به عنوان یک الگوریتم نهان‌نگاری مقاوم[3] و یا یک الگوریتم نهان‌نگاری شکننده[4] اسـتفاده نمود. در واقع برای محاسبه هر‌یک از بردارهای حـوزه ICA تصویر نهان‌نگاری‌شده، r مؤلفه با انرژی کمتر از بردار حوزه ICA متناظر از تصویر اصلی، با r مؤلفه با انرژی بیشتر از بردار حوزه ICA متناظر از واترمارک جایگزین می شود. مشابه تبدیلات فرکانسی، در اینجا نیز مؤلفه‌های اولیه (مؤلفه‌هایی که انرژی بیشتری دارند ) سازنده کلیات تصویر و مؤلفه‌های با انرژی کمتر، سازنده جزئیات و لبه‌های تصویر می‌باشند. در اینجا، r متغیری است که بین مقاومت الگوریتم نهان- نگـاری و میـزان رؤیت‌پـذیری واترمـارک ( اعوجـاج ناشـی از جاسازی واترمارک ) موازنه برقرار می‌نماید.

الگوریتم S.Bounkong [۱۶]، تلاشی به جهت ارائه یک الگوریتم با کیفیت بهتر از الگوریتم [۱۵] می‌باشد. در اینجا نیز، ابتدائاً تصویر به بلوکهای ناهمپوش تقسیم شده و به حوزه ICA برده شـده اسـت. بلوکهای تعریف شده در این الگوریتم، ۱۶×۱۶ می‌باشند. در ادامـه، دسته خاصی از ضرایب حوزه ICA هر بلوک انتخاب گردیـده و بـا استفاده از رابطه:

$$m = 0 \quad x \to \Delta \times \lfloor x/\Delta \rfloor$$
$$m = 1 \quad x \to \Delta \times \lfloor x/\Delta \rfloor + \Delta/2$$
(۱۳)

کوانتیزه گردیـده اسـت. در ایـن رابطـه، m، بیتـی از رشـته بیـت واترمارک، x ، مقدار ضریب ، $\lfloor \ \rfloor$ ، اپراتور جـزء صـحیح و $\Delta$، گـام کوانتیزاسیون می‌باشد. واضح است کـه هرچـه گـام کوانتیزاسـیون بزرگتر باشد، مقاومت بیشتری در برابر حملات خواهیم داشت و در عوض شاهد تخریب بیشتری در تصویر نهان‌نگاری شده خواهیم بود. در انتخاب گام کوانتیزاسیون باید موازنه‌ای بین این دو فاکتور برقرار گردد. گام کوانتیزاسیون در این الگوریتم، به‌گونه‌ای انتخاب شده است که PSNR بیش از 44dB را برای تصویر نهان‌نگاری‌شده بـه همـراه آورد. بـرای بازیـابی واترمـارک نیـز در حالـت کلـی، دو آشکارسـاز نزدیکترین همسایه و آشکارساز MAP پیشنهاد شده و عملکرد آنها مقایسه شده است. در آشکارسـاز نزدیکتـرین همسـایه، بـر اساس نزدیکی بیشتر $\lfloor x/\Delta \rfloor \times \Delta - x$ به صفر و یا $\Delta/2$ ، در مـورد صـفر و یـا یک بودن بیت واترمارک مربوطه تصمیم‌گیری می‌شود. در عوض،

---

[3] Robust Watermarking
[4] Fragile Watermarking

[1] Visual Cortex
[2] Human Visual System

# ICA و کاربرد آن در نهان نگاری تصاویر دیجیتال

آشکارساز MAP، واترمارک جاسازی شده (m) را بگونه‌ای تخمین می‌زند که احتمال پسین $p(m|\tilde{s})$، بیشینه مقدار ممکن گردد. در این رابطه، $\tilde{s}$ همان تصویر نهان‌نگاری شده (در حوزه ICA) می‌باشد. با استفاده از قانون بیز می‌توان نوشت:

$$p(m|\tilde{s}) = \frac{p(\tilde{s}|m)p(m)}{p(\tilde{s})} \quad (14)$$

با توجه به اینکه می‌توان m را متغیری با توزیع یکنواخت فرض نمود و با توجه به اینکه $p(\tilde{s})$ هیچ وابستگی خاصی به m ندارد، بیت iام از واترمارک را می‌توان بصورت زیر تخمین زد:

$$\hat{m}_i \propto \max_m p(\tilde{s}_i | m_i) \quad (15)$$

که در آن هر $m_i$ می‌تواند مقدار صفر یا یک داشته باشد. همانطور که از این معادله نیز برمی‌آید، بهبود کیفیت آشکارساز درگرو درک بهتر از حملات اعمال شده بر تصویر نهان‌نگاری شده می‌باشد. تفاوت الگوریتم‌های مختلف ارائه شده در [20]، در نحوه انتخاب ضرایبی که جاسازی واترمارک در آنها صورت می‌گیرد می‌باشد. در ICA-Sel، در هر بلوک، از میان IC های شماره 5، 55 و 59 که بر اساس مشاهدات، مقاومت بیشتری در برابر حملات نویز گاوسی، فشرده سازی JPEG و فشرده سازی SPIHT دارند، یکی بطور تصادفی انتخاب شده و بیت واترمارک در آن جاسازی گردیده و از آشکارساز نزدیکترین همسایه، به جهت بازیابی واترمارک استفاده می‌گردد. در ICA-Ne و ICA-Map، از IC که مطابقت بیشتری با مدل تصویر اصلی و مدل حملات اعمال شده به آن دارد، استفاده می‌شود. بر این اساس، از تست $\chi^2$ که یک تست آماری مشخص‌کننده تبعیت داده‌های واقعی از یک مدل تصادفی استفاده شده است. در اینجا، از میان IC های موجود، IC به عنوان حوزه جاسازی واترمارک انتخاب می‌شود که مقدار $\chi^2$ مربوط به مدل تصویر اصلی و نیز مدل حمله آن، کمترین مقدار در بین مقادیر معادل مربوط به ICهای دیگر می‌باشد. تفاوت این دو روش، در نحوه آشکارسازی واترمارک می‌باشد.

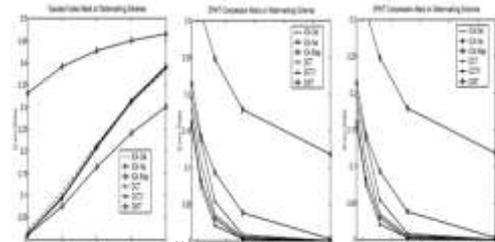

**شکل 3) عملکرد الگوریتم S.Bounkong در برابر حملات 1) نویز گاوسی 2) فشرده سازی JPEG 4) فشرده سازی SPIHT [16]**

الگوریتم ارائه شده توسط W.Lu [18]، الگوریتم‌های ارائه شده توسط J.Murillo [20-22]، الگوریتم ارائه شده توسط F.Gui [23] و الگوریتم ارائه شده توسط J.Liu [24]، دیگر الگوریتم‌های نهان‌نگاری می‌باشند که در آن از جاسازی واترمارک در حوزه ICA تصویر استفاده شده است.

## 5- بازیابی واترمارک

بحث شد که ICA یک روش جداسازی کور منابع می‌باشد و می‌تواند تحت شرایط خاصی، مخلوط چند سیگنال را به سیگنالهای تشکیل‌دهنده‌اش تجزیه نماید. درباره شرایط خاصی که مجاز به استفاده از تبدیل می‌باشیم و محدودیتها و ابهامات خروجی الگوریتم نیز مفصلا بحث شد. در این بخش از مقاله به بررسی الگوریتم‌های نهان‌نگاری پرداخته شده است که از ICA در بلوک بازیابی واترمارک خود استفاده می‌کنند. ایده اصلی در اینجا، امکان درنظرگرفتن تصویر نهان‌نگاری شده یا حوزه تبدیل آن، بصورت ترکیبی خطی از تصویر اصلی، کلید و واترمارک می‌باشد. بلوک جاسازی واترمارک در این الگوریتم‌ها، محقق‌کننده این فرض می‌باشد. حال اگر به نوعی بتوان سه نسخه غیر وابسته از تصویر نهان‌نگاری‌شده را در ورودی بلوک بازیابی واترمارک تولید نمود، می‌توان از تحلیل ICA برای بدست‌آوردن این سه سیگنال مستقل تشکیل‌دهنده تصویر نهان‌نگاری‌شده؛ یعنی تصویر اصلی، واترمارک و کلید؛ استفاده نمود. نکته بسیار ظریف و مهمی که در این میان وجود دارد این است که تولید این سه تصویر ترکیبی باید صرفا برای کاربران مجاز، ممکن باشد و فردی که کلید را در اختیار ندارد، نباید بتواند چنین تصویرهایی تولید نماید. عدم توجه به این نکته، امنیت الگوریتم نهان‌نگاری را با چالشی جدی روبرو می‌نماید. چرا که هرکس می‌تواند با استفاده از تبدیل ICA، تصویر اصلی و واترمارک را بدست آورد.

الگوریتم Yu-Sattar [25] نخستین الگوریتمی است که در آن از ICA در بلوک بازیابی واترمارک جاسازی شده در حوزه مکان، استفاده شده است. علیرغم استفاده از ICA در بلوک بازیابی واترمارک، این الگوریتم، یک الگوریتم غیرکور می باشد. WMicaD [26]، الگوریتم دیگری از این خانواده می‌باشد. مهمترین مزیت این الگوریتم نسبت به [25]، کور بودن آن می باشد.

در [27-29] از ICA برای بازیابی واترمارک جاسازی شده در حوزه DWT[1] استفاده شده است. برای بازیابی واترمارک در این الگوریتم‌ها، ابتدا سیگنال به حوزه موجک برده می‌شود و سپس ICA اعمال می‌گردد. به عبارت دیگر، ورودی ICA در این الگوریتم‌ها، از ضرایب حوزه موجک تولید می شود.

یک مشکل اساسی الگوریتم‌های حوزه DWT این است که ابتدا باید

---

[1] Discrete Wavelet Transform

سیگنال به حوزه موجک (DWT) برده شود و سپس از ICA استفاده شود. به جهت حذف این مرحله، در [۳۰-۳۲] از جاسازی در حوزه RDWT[1] استفاده شده است. با این کار، مقاومت الگوریتم در برابر حملات نیز افزایش پیدا کرده است.

از ICA برای بازیابی واترمارک جاسازی شده در دیگر حوزه‌ها نیز استفاده شده است. در [۳۳] از ICA برای بازیابی واترمارک جاسازی شده در حوزه SVD[2] و در [۳۴]، از آن برای بازیابی واترمارک جاسازی شده در حوزه Contourlet استفاده شده است. روش کار در همه الگوریتم‌های استفاده‌کننده از ICA در بلوک بازیابی واترمارک، تقریبا مشابه هم می‌باشد. ساختار کلی بلوک جاسازی واترمارک الگوریتم Hien [۲۸] در شکل ۴ و ساختار کلی بلوک بازیابی واترمارک این الگوریتم در شکل ۵ نشان داده شده است. واترمارک در این الگوریتم، یک رشته بیت باینری می‌باشد که از درهم ریزی مطابق کلید یک Logo حاصل می‌شود. این واترمارک، در زیرباندهای LH3 و HL3 تصویر اصلی جاسازی می شود. برای جاسازی واترمارک در این الگوریتم، از ضرایب LH3 و HL3 تصویر اصلی استفاده می‌شود. اگر تصویر اصلی m×n باشد، ماتریس $I_l^{LH}$ حاوی ضرایب LH3 تصویر اصلی و ماتریس $I_l^{HL}$ حاوی ضرایب HL3 تصویر اصلی، ماتریس‌هایی با ابعاد (m/8)×(n/8) خواهند بود. واضح است که واترمارک (W) نیز حداکثر می‌تواند، چنین ابعادی داشته‌باشد. جاسازی واترمارک در این الگوریتم، یک جاسازی مبتنی بر محتوا می‌باشد. رابطه جاسازی واترمارک در این الگوریتم، بصورت:

$$I_l^{'LH}(i,j) = I_l^{LH}(i,j)$$
$$+ A^{LH}.\alpha(1-NVF(i,j))W(i,j)$$
$$+ B^{LH}.\beta.NVF(i,j)W(i,j) \quad (16)$$
$$I_l^{'HL}(i,j) = I_l^{HL}(i,j)$$
$$+ A^{HL}.\alpha(1-NVF(i,j))W(i,j)$$
$$+ B^{HL}.\beta.NVF(i,j)W(i,j)$$

می‌باشد. در این رابطه، **α** و **β** ، تنظیم کننده میزان ناپیدایی واترمارک می باشند. برای **β** ، مقدار ثابت ۰/۱ درنظر گرفته شده است اما **α** ، متناسب با شرایط، تنظیم می شود. در این رابطه، '$A^{LH} = \mu | I_l^{LH} |$' ، '$A^{HL} = \mu | I_l^{HL} |$' و $B^{HL} = A^{HL}/10$ و $B^{LH} = A^{LH}/10$ می‌باشد. جاگذاری ماتریس‌های $I_l^{'LH}$ و $I_l^{'HL}$ در تصویر اصلی و اعمال IDWT ، ما را به تصویر نهان‌نگاری شده می‌رساند . کلید بازیابی واترمارک

(Key) نیز با استفاده از روابط زیر بدست می آید: (17)
$$I_l^{*LH}(i,j) = I_l^{LH}(i,j) + A^{LH}.\alpha_k.K(i,j)$$
$$I_l^{*HL}(i,j) = I_l^{HL}(i,j) + A^{HL}.\alpha_k.K(i,j)$$
$$Key(i,j) = I_l^{*LH}(i,j) + I_l^{*HL}(i,j) + K(i,j)$$

که در آن، K، ماتریس کلید بوده و $\alpha_k = 0.5$ می باشد. دو کلید K و Key باید بصورت امنی به گیرنده منتقل شود .

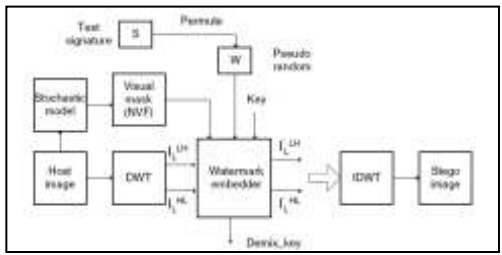

**شکل ۴) بلوک جاسازی واترمارک در الگوریتم Hien [۲۸]**

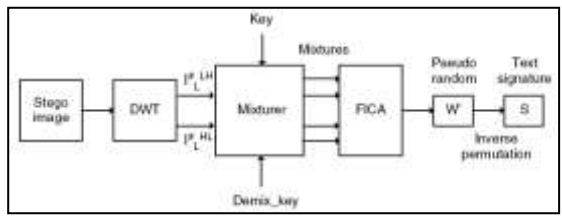

**شکل ۵) بلوک بازیابی واترمارک در الگوریتم Hien [۲۸]**

برای بازیابی واترمارک در این الگوریتم، ابتدا چهار ماتریس ترکیبی :

$$X_1 = I_l^{\#LH} + Key$$
$$X_2 = I_l^{\#HL} + Key$$
$$X_3 = I_l^{\#LH} + I_l^{\#HL} \quad (17)$$
$$X_4 = Key + K$$

محاسبه می شود. $I_l^{\#HL}$ و $I_l^{\#LH}$ در روابط فوق، به ترتیب، ضرایب HL3 و LH3 تصویر دریافتی می‌باشند. واضح است که هر یک از ماتریس‌های $X_1$ ، $X_2$ ، $X_3$ ، $X_4$ را می‌توان بصورت ترکیبی از K، W، $I_l^{HL}$ و $I_l^{LH}$ نوشت. بنابراین با اعمال ICA بر این ماتریس‌های ترکیبی، می توان واترمارک W و نهایتا رشته بیت پیام را تعیین نمود.

در [۳۲]، جاسازی واترمارک در حوزه RDWT می‌باشد. تنها تفاوت موجود بین DWT و RDWT این است که در RDWT، پس از فیلتر کردن، کاهش نرخ نمونه‌برداری صورت نمی گیرد. به عبارت دیگر، در اینجا شاهد برابری ابعاد زیرباندها با ابعاد تصویر اصلی می‌باشیم. عدم کاهش نرخ نمونه‌برداری در این تبدیل، Shift Invariant بودن و مقاومت بیشتر در برابر نویز جمع شونده را به

---
[1] Redundant Wavelet Transform
[2] Singular Value Decomposition

ارمغان می آورد. استفاده از این حوزه، سبب تسهیل پروسه بازیابی واترمارک در الگوریتم Hien نیـز مـی‌باشـد. خطـی بـودن تبـدیل RDWT و برابری ابعاد زیرباندها با ابعاد تصویر اصلی در این تبدیل، سـبب حـذف مرحلـه بـه حـوزه تبـدیل بـردن، در پروسـه بازیـابی واترمارک الگوریتم می‌گردد. برخلاف الگوریتم‌های دیگر، در اینجـا، الگوریتم ICA پروسه بازیابی واترمارک، روی خود تصویر نهان‌نگاری شده و نه حوزه تبدیل آن اجرا می شود که این امر سبب کاهش بار محاسباتی پروسه بازیابی واترمارک می گردد.

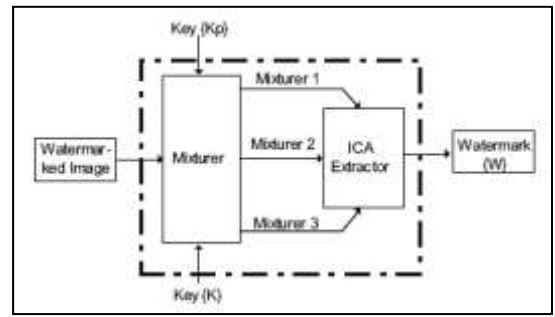

**شکل ۶) بلوک بازیابی واترمارک الگوریتم RDWT-Hien [۳۲]**

## ۶- ابزار حمله

در حال حاضر، مهمترین و پرچـالش‌تـرین کـاربرد ICA در عرصـه نهان‌نگاری، کاربرد آن به عنوان ابزار حمله می‌باشد. این امـر از آن رو است که حملات امنیتی جدیدترین و ویران کننده‌ترین حملاتی هسـتند کـه تـاکنون مطـرح شـده‌انـد و ICA یکـی از متـداول‌ترین ابزارهای اعمال این حملات می باشد. نکته اصلی که حملات مبتنی بر ICA از آن استفاده می نمایند، استقلال سیگنال واترمارک از سیگنال حامل می باشد.

بیشتر کارهای انجام شده در دهـه نـود مـیلادی کـه دهـه آغـازین تحقیقات در زمینـه نهان‌نگـاری مـی باشـد، راجـع بـه مقاومـت[1] الگوریتم‌ها می باشد. در این دهه، اصلا تفـاوتی بـین مقاومـت و امنیت در الگوریتم‌های نهان‌نگاری، وجود ندارد و ایجاد مقاومـت در برابر حملات کوری مانند نویز جمع شـونده، فشرده‌سـازی، بـرش، چرخش و امثالهم که بعدا از آنها تعبیر به حملات غیرعمـدی[2] شـد [۳۵]، در کانون توجـه قـرار دارد. در ادامـه، حمـلات هوشـمندتری مطرح شدند که در آنها از اطلاعات موجود در مورد الگوریتم نهان‌نگاری، استفاده می‌شـد. حملـه تبـانی[3] [۳۶] و حملـه حساسـیت[4]

[۳۸-۳۶]، از جمله این حملات می‌باشند. طرح این حملات، نیـاز به تفکیک بیشتر بین دو مقوله امنیت و مقاومت و پرداخت جدی تر به مقوله امنیت در الگوریتم‌های نهان‌نگاری را روشن ساخت [۳۹]. همچنین مشخص شد که عمدی یا غیرعمدی بودن یک حمله نمی‌تواند ملاک تفکیک خوبی باشد و بایـد از ملاکهای دیگـری بـرای تفکیک این دو مفهوم بسیار نزدیک به هم، استفاده نمـود. چـرا کـه فرضا با استفاده عمدی از فشرده سازی که یک حملـه غیر عمـدی می‌باشد نیز ممکن است بتـوان واترمارک جاسازی شـده در یـک تصویر نهان‌نگاری شده خاص را حذف نمود.

در [۴۰] برای اولین بار به طبقه‌بندی حملات تهدید کننـده امنیـت الگوریتم نهان‌نگاری، پرداخته شده است. این طبقـه‌بنـدی، مشـابه دسته‌بندی اسـت کـه Diffie-Hellman در مـورد الگوریتم‌هـای رمزنگاری ارائه داده اند. بر این اساس، حملات به سه دسته حمـلات WOA[5]، KMA[6] و KOA[7] تقسـیم شـده‌انـد. در حـال حاضـر، روشهای BSS[8] مانند ICA، اصلی‌ترین روش اعمال حملات امنیتی امنیتی می‌باشد. در حمـلات WOA، تنهـا تعـداد $N_0$ بـردار نهان‌نگاری‌شده در اختیار است. این حملات، خطرناک‌ترین نوع حملات می‌باشند. چرا که در آن، مهاجم بـه اطلاعـاتی بجـز تصاویر نهان‌نگاری‌شده نیاز ندارد. در حملات KMA، مهاجم به $N_0$ بردار نهان‌نگاری شده و پیام جاسازی شده در آنها دسترسی دارد و در حملات KOA، مهاجم به $N_0$ بردار نهان‌نگاری‌شده و معادل آنها از تصویر اصلی دسترسی دارد.

در [۴۰]، از ICA به جهت ایراد یک حملـه WOA بـه الگوریتم‌های نهان‌نگاری طیف گسترده استفاده شده است. در حالـت کلـی، پروسه جاسازی واترمارک در الگوریتم‌های نهان‌نگاری طیف گسترده را می‌توان با استفاده از رابطه:

$$Y = X + US_m \qquad (۱۸)$$

نمایش داد. $X$ در این رابطه، ماتریسی مـی باشـد کـه سـتون‌های آن بردارهای ویژگی استخراج شـده از تصـویر اصـلی مـی‌باشـد. بسـتر جاسازی واترمارک، همین بردارهای ویژگی می‌باشند. ماتریس $Y$، ماتریس در برگیرنده بردارهای نهان‌نگاری شده می‌باشد. هر ستون از این ماتریس، معادل یک ستون از ماتریس $X$ می‌باشد و دلالت بـر یک بردار نهان‌نگاری‌شده دارد. اگـر $N_v$، طـول بـردار ویژگـی و $N_0$، تعداد بردارهای نهان‌نگاری‌شده موجود باشد، ماتریس‌های $X$ و $Y$،

---

[4] Sensitivity Attack
[5] Watermarked Only Attack
[6] Known Message Attack
[7] Known Original Attack
[8] Blind Source Separation

[1] Robustness
[2] Non-Intentional Attack
[3] Collusion Attack

# ICA و کاربرد آن در نهان نگاری تصاویر دیجیتال

ماتریس‌هایی به ابعاد $N_v \times N_0$ خواهند بود. $U$ در این رابطه، ماتریسی به ابعاد $N_v \times N_c$ می‌باشد که از چینش سـطری بردارهـای گسـترش دهنده $\bar{u}_i$، که بردارهایی ستونی می‌باشند، تولید شده است. $N_c$ در اینجا مبین تعداد بیت پیام جاسازی شده در هر بردار ویژگی مـی‌ باشد. $S_m$ در این رابطه نیز ماتریسی به ابعاد $N_c \times N_0$ می باشد کـه ستون‌های آن، رابطه مستقیمی با پیام جاسازی شـده در هـر بـردار ویژگـی دارد. فرضـا در الگـوریتم نهـان‌نگـاری طیـف گسـترده کلاسیک[۴۱]، داریم:

$$\bar{y} = \bar{x} + \sum_{i=1}^{N_c} b(i)\bar{u}_i \qquad (19)$$

که در آن، $\bar{x}$، بردار استخراج شده از تصویر اصلی، $\bar{y}$، بردار نهان‌-نگاری شده معادل، $\bar{b}$، رشـته $N_c$ بیتـی پیـام ($\bar{b} \in \{-1,1\}^{N_c}$) و $\bar{u}_i$ ($i=1,...,N_c$)، بردارهـای $N_v$ بیتـی گسترش‌دهنـده مـی‌باشـد (بردارهای $\bar{u}_i$، انرژی هر بیت پیام را بر روی کل بـردار $\bar{x}$ - بطـول $N_v$ پخش می‌نماید). رابطه کلی در الگوریتم نهـان‌نگـاری طیـف گسترده پیشرفته [۳۸] نیز فرمی بصورت:

$$\bar{y} = \bar{x} + \sum_{i=1}^{N_c}(\alpha b(i) - \lambda \frac{z_{\bar{x},\bar{u}_i}}{||\bar{u}_i||})\bar{u}_i \qquad (20)$$

دارد که در آن :

$$z_{\bar{x},\bar{u}_i} = \sum_{k=1}^{N_v} \bar{x}(k)\bar{u}_i(k) \qquad (21)$$

بوده و $\alpha$ و $\lambda$، پارامترهای تنظیم کننده میزان ناپیدایی واترمارک و مقاومت آن می باشند.

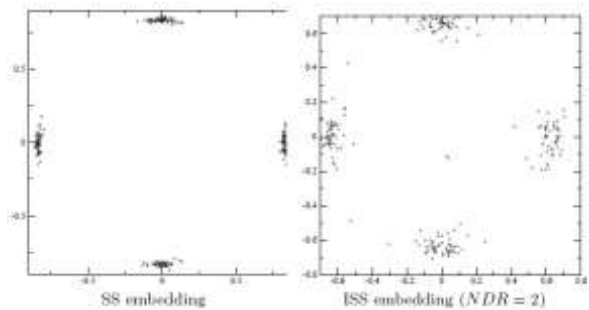

شکل ۶) همبستگی نرمالیزه بین مقدار اصلی بردار گسترش دهنده و مقادیر تخمینی توسط الگوریتم ($N_0=1000$، WCR=-21dB، $N_v=512$)

حال اگر بتـوانیم $X$ در رابطـه $Y = X + US_m$ را بـه عنـوان نـویز درنظر بگیریم و اگر فرض استقلال بیت‌های پیام جاسازی شده در هر بردار ویژگی، فرض درستی باشد و اگر $N_0$ بـردار نهـان‌نگـاری‌شـده موجـود بـوده و اگـر از کلیـد مشـابهی بـرای جاسـازی واترمـارک در بردارهای ویژگی استفاده شده باشد، به راحتی می‌توان بـا اسـتفاده از تجزیـه ICA، بیت‌هـای واترمـارک جاسـازی شـده در هـر بلـوک و بردارهای گسترش‌دهنده را تعیین نمـود. ایـن فرضـیات، فرضـیات چندان دور از ذهنی نمی‌باشند.

شکل ۶، موفقیت‌آمیز بودن حمله در حالتی که ۱۰۰۰ تصویر نهان نگاری شده با WCR برابر 21dB- حاوی دو بیت واترمارک جاسازی شده ( با استفاده از بردارهای بسط‌دهنده ۵۱۲ بیتی) موجـود مـی‌ باشد را نشان می‌دهد. انتخاب $N_c=2$ در اینجـا، صرفـا بـه جهـت سهولت نمایش موفقیت حمله می‌باشد. شکل ۶، نتیجه اجرای حمله در ۱۰۰ آزمایش متفاوت می‌باشد. خروجی هر آزمایش، تخمینـی از دو بردار بسط‌دهنده ۵۱۲ بیتی استفاده‌شده مـی‌باشـد. هـر نقطـه ($c_1,c_2$) در شکل ۶، از محاسبه همبستگی نرمالیزه بین این دو بردار و دو بردار بسط دهنده استفاده‌شده در پروسه جاسازی واترمـارک، بدست آمده است. واضح است که در حالت کلی، هرچه تعداد نقاط نزدیک به نقاط (0,1)، (0,-1)، (1,0) و (-1,0) بیشتر باشد، حمله صورت گرفته از توفیق بیشتری برخوردار بوده است. همانطور که در شکل مشاهده مـی‌شـود، عملکـرد حملـه در الگـوریتم SS، بهتـر از الگوریتم ISS می‌باشد. این امر از آن سو است که واریانس واترمـارک جاسازی شده به روش ISS، کمتر از واترمـارک جاسـازی شـده بـه روش SS می باشد.

به راحتی می‌توان از ICA به جهت اعمال حمله KOA به الگوریتم‌-های نهان‌نگاری طیف‌گسترده نیز استفاده نمود. در حالت کلی، اگر بیت‌های پیام جاسازی شده در هر بردار ویژگی، مستقل از هم باشند، با فرض اینکه از کلید مشابهی برای جاسازی واترمارک در بردارهای ویژگی استفاده شده باشد، با استفاده از تجزیه ICA می‌توان بیت‌های واترمارک جاسازی شده در هر بلوک و بردارهای گسـترش‌دهنـده را تعیین نمود. برخلاف قبل که مـاتریس $X$ بـه عنـوان نـویز در مـدل ICA در نظر گرفته می‌شد، مدل ICA که در اینجـا اسـتفاده مـی‌ شود، مدل بدون نویز می‌باشد:

$$D = Y - X = US_m \qquad (22)$$

بدیهی است که میزان توفیق این حمله در تخمین صحیح بردارهای گسترش‌دهنده، بسیار بیشتر از حملات WOA بوده و این حمله در کل، حمله ساده‌تری می‌باشد. البته، مشابه قبل، تخمین مطروحه در اینجا نیز دارای ابهام علامت می‌باشد. به عبارت دیگر، آنچه تخمـین زده می‌شود، ممکن است خود بردار گسترش‌دهنده بکار رفته در جاسازی واترمارک ($\bar{v}$) و یا قرینه آن ($-\bar{v}$) باشـد. بـه تبـع آن، واترمارک تخمین زده شده در این حمله نیز دارای ابهام علامـت باشد. همانطور که در بخش دوم مقاله نیز بحث شد، این امـر از ماهیت ICA نشأت می‌گیرد.

گزارشاتی در مورد استفاده از ICA به جهت ایراد حمله بـه دیگـر الگوریتم‌های نهان‌نگاری نیز موجود می‌باشد. فرضا در [۴۲] از ICA به جهت ایراد حمله به الگوریتم‌های مبتنی بر DM [۴۳] و تخمـین واترمارک جاسازی شده در سیگنال حامل، استفاده شده است. اصول حمله در اینجا، مشابه قبل می‌باشد.

## ۷- نتیجه‌گیری

در این مقاله پس از تبیین تکنیک پردازش سیگنال ICA و ویژگیها و محدودیتهای آن، به تشریح کاربردهای سه‌گانه آن در عرصه نهان‌-نگاری تصاویر دیجیتال پرداخته شد. در باب قابلیت استفاده از حوزه ICA به عنوان حوزه جاسازی واترمارک و ویژگیهایی کـه نتیجـه مستقیم شباهت مشخصات بردارهای پایه حوزه ICA و سلولهای اولیه و ثانویه پوسته بینایی در سیستم بینایی انسان می‌باشد، صحبت شد. در ادامه نشان داده شد که با طراحی مناسب می‌تـوان الگوریتم‌هایی طراحی نمود که از ICA در بلوک بازیابی واترمارک خود استفاده نمایند و نهایتا بحث شد که ICA را می‌تـوان یکـی از اصلی‌ترین ابزارهای نقض امنیتی الگوریتم‌های نهان‌نگاری و استگانو-گرافی دانست. طرح ICA به عنوان ابزار حمله، الزامـات تئوریـک و عملـی جدیـدی را در طراحـی الگـوریتم‌هـای نهـان‌نگـاری و استگانوگرافی مقاوم و امن ایجاب نموده است.

## ۸- مراجع

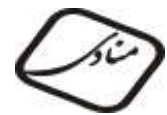

# رویدادها 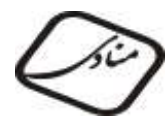

- **دانشگاه صنعتی خواجه نصیر الدین طوسی**

  - کارگاه علمی رمز نگاری در روزهـای پنجشــنبه و جمعــه ۱۴ و ۱۵ آبان۱۳۸۸ برگزار شد.

  دکتر احمدیان
  نماینده کمیته علمی انجمن رمز ایران
  در دانشگاه صنعتی خواجه نصیرالدین طوسی

- **دانشگاه اصفهان**

  کارگاه آموزشی امنیت شبکه به شرح زیر در دانشگاه اصفهان برگزار گردید :

  - پویش شبکه (محمدعلی طائبی، حسین سلیمانی) آذر ماه ۸۸
  - مجوز سطح پروسه (امیرمهدی قاسمی) آذر ماه ۸۸.
  - سمینار کلیک ربایی توسط: سید محمدحسین میرشاه جعفر اصفهانی آذر ۸۸.

  دکتر بهروز ترک لادانی
  نماینده کمیته علمی انجمن رمز ایران
  در دانشگاه اصفهان

- **دانشگاه صنعتی اصفهان**

  - در تاریخ مهر ۱۳۸۸، حمله تمایز علیه الگوریتم‌های دنباله‌ای توسط آقای مهندس آرش میرزایی زیر نظر دکتر دخیل علیان ارایه شد.

  دکتر محمد دخیل‌علیان
  نماینده کمیته علمی انجمن رمز ایران
  در دانشگاه صنعتی اصفهان

- **شهید بهشتی**

  - دفاع از پایان نامه "سیستم پرسش و پاسخ دامنه باز" در تاریخ ۸۸/۸/۲۷ توسط احد نیک نیا برگزار گردید. اسـتاد راهنمـا: دکتـر لیلا شریف، استاد مشاور : دکتر زیبا اسلامی.

  دکترحسین حاجی ابوالحسن
  نماینده کمیته علمی انجمن رمز ایران
  دانشگاه شهید بهشتی

- **دانشگاه صنعتی شریف**

  - ارئه سمینار طرحهای تسهیم چند مرحله‌ای چند راز توسط خانم فاطمی در تاریخ آذر ماه ۸۸ انجام شد.

  دکتر ترانه اقلیدس
  نماینده کمیته علمی انجمن رمز ایران
  در دانشگاه صنعتی شریف